\documentclass[prl,aps,superscriptaddress,floatfix,tightenlines,showpacs,twocolumn]{revtex4-1}
\usepackage{graphicx}
\usepackage{dcolumn}   
\usepackage{bm}        
\usepackage{amssymb}   
\usepackage{amsmath}
\usepackage{url}
\usepackage{layout}
\usepackage{color}
\usepackage{epsfig}
\usepackage{graphicx}
\usepackage{mathrsfs}\usepackage{bm}\usepackage{appendix}\usepackage{color}
\usepackage{makecell}
\usepackage{booktabs}
\usepackage{float}
\usepackage[hyperindex,breaklinks]{hyperref}

\def\be{\begin{equation}}       \def\ee{\end{equation}}
\def\bea{\begin{eqnarray}}      \def\eea{\end{eqnarray}}
\def\bp{\begin{pmatrix}} \def\ep{\end{pmatrix}}
\def\beaa{\begin{equation}\begin{aligned}}
		\def\eeaa{\end{aligned}\end{equation}}

\begin{document}

\title{Interlayer Coupling Driven
High-Temperature Superconductivity in La$_3$Ni$_2$O$_7$ Under Pressure}

\author{Chen Lu}
\thanks{These two authors contributed equally to this work.}
\affiliation{New Cornerstone Science Laboratory, Department of Physics, School of Science, Westlake University, Hangzhou 310024, Zhejiang, China}
\author{Zhiming Pan}
\thanks{These two authors contributed equally to this work.}
\affiliation{Institute for Theoretical Sciences, Westlake University, Hangzhou 310024, Zhejiang, China}
\affiliation{New Cornerstone Science Laboratory, Department of Physics, School of Science, Westlake University, Hangzhou 310024, Zhejiang, China}
\author{Fan Yang}
\email{yangfan\_blg@bit.edu.cn}
\affiliation{School of Physics, Beijing Institute of Technology, Beijing 100081, China}
\author{Congjun Wu}
\email{wucongjun@westlake.edu.cn}
\affiliation{New Cornerstone Science Laboratory, Department of Physics, School of Science, Westlake University, Hangzhou 310024, Zhejiang, China}
\affiliation{Institute for Theoretical Sciences, Westlake University, Hangzhou 310024, Zhejiang, China}
\affiliation{Key Laboratory for Quantum Materials of Zhejiang Province, School of Science, Westlake University, Hangzhou 310024, Zhejiang, China}
\affiliation{Institute of Natural Sciences, Westlake Institute for Advanced Study, Hangzhou 310024, Zhejiang, China}

\begin{abstract}
The newly discovered high-temperature superconductivity in La$_3$Ni$_2$O$_7$ under pressure has attracted a great deal of
attentions.
The essential ingredient characterizing the electronic properties is the bilayer NiO$_2$ planes
coupled by the interlayer bonding of $3d_{z^2}$ orbitals through the intermediate oxygen-atoms.
In the strong coupling limit, the low energy
physics is described
by an intralayer antiferromagnetic spin-exchange interaction $J_{\parallel}$ between $3d_{x^2-y^2}$ orbitals and an interlayer one $J_{\perp}$ between $3d_{z^2}$ orbitals.
Taking into account Hund's rule on each site and integrating out the $3d_{z^2}$ spin degree of freedom, the system reduces to a single-orbital bilayer $t$-$J$ model
based on the $3d_{x^2-y^2}$ orbital.
By employing the slave-boson approach, the self-consistent equations for the bonding and pairing order parameters are solved.
Near the physically relevant $\frac{1}{4}$-filling regime (doping $\delta=0.3\sim 0.5$),
the interlayer coupling $J_{\perp}$ tunes the conventional single-layer $d$-wave superconducting state to the $s$-wave one.
A strong $J_{\perp}$ could enhance the inter-layer superconducting order, leading to a dramatically increased $T_c$.
Interestingly, there could exist a finite regime in which an $s+id$ state emerges.
\end{abstract}
\maketitle

Since the discovery of the high-temperature superconductivity (SC) in cuprates \cite{bednorz1986LBCO}, understanding the pairing mechanism of unconventional SC \cite{anderson1987rvb,kotliar1988,lee2006htsc,keimer2015highTc,proust2019highTc} and searching for new superconductors with high critical temperature $T_c$ remain long-term
challenges.
It has been widely believed that strong electron correlations drive
the $d$-wave pairing symmetry in  high-$T_c$ superconductors \cite{anderson1987rvb,kotliar1988,lee2006htsc}.
Under such an understanding, many attempts have been made to search for high-$T_c$ superconductors in materials with strong electron correlations,
especially, the 3$d$-transition metal oxides\cite{anisimov1999nickelate,li2019nickelate,HuLH2019,zhang2020ni,botana2021nickelate,zeng2022nickelate,LuC2022}.
However, no new family of superconductors have been
synthesized with $T_c$ above the nitrogen boiling point  until the recent discovery of La$_3$Ni$_2$O$_7$ (LNO)~\cite{Wang2023LNO}.
It exhibits the superconducting $T_c \approx 80$ K under pressures over $14$GPa , which
has attracted considerable attentions both experimental \cite{WenHH2023,Wang2023LNOb,YuanHQ2023LNO} and theoretical \cite{YaoDX2023,Dagotto2023,WangQH2023,lechermann2023,Kuroki2023,HuJP2023,ZhangGM2023DMRG,Werner2023,shilenko2023correlated,WuWei2023charge,cao2023flat,chen2023critical,YangF2023,oh2023type2,qu2023bilayer,Yi_Feng2023}.

Similarly to cuprates, LNO hosts a layered structure \cite{Wang2023LNO,Wang2023LNOb,WenHH2023} with each unit cell containing two conducting NiO$_2$ layers, which is isostructural with the CuO$_2$ layer in cuprates. Calculations based on the density-functional-theory (DFT) \cite{pardo2011dft,Wang2023LNO} suggest that the low-energy degrees of freedom near the Fermi level are of the Ni-$3d$ orbitals, including two $E_g$-orbitals, {\it i.e.}, $3d_{z^2}$ and $3d_{x^2-y^2}$, and the site energy of the former is lower than that of the latter.
The four $E_g$-orbitals in two Ni$^{2.5+}$ cations within a unit cell share three electrons in total.
The $3d_{z^2}$ orbitals in the two layers within a
unit cell couple via the hybridization with the O-$2p$
orbitals in the intercalated LaO layer.
Under pressure, such a Ni-O-Ni bonding angle along the $c$-axis changes from $168^\circ$ to $180^\circ$~\cite{Wang2023LNO}, which
greatly enhances the effective interlayer coupling, as suggested by the combination of the synchrotron X-ray diffraction and DFT calculations~\cite{Wang2023LNO}.
The high-$T_c$ SC only emerges under pressure, implying that the interlayer coupling is crucial for the high-$T_c$ SC in LNO.

The $3d$-orbital character of the low-energy degrees of freedom in LNO suggests strong electron correlations.
Such a viewpoint is supported by a recent experiment\cite{WenHH2023}
which reveals that LNO is in the proximity of Mott phase and exhibits non-Fermi-liquid behavior.
Therefore, the strong-coupling picture should be legitimate for exploring the pairing mechanism therein.
It has been proposed in Ref.~\cite{ZhangGM2023DMRG,Yi_Feng2023} that the interlayer coupling between the two Ni-$3d_{z^2}$ orbitals along the rung
within a unit cell would induce the antiferromagnetic (AFM) super exchange interaction. 
The same viewpoint is adopted here. However, an important ingredient has been missed in these studies,
i.e., the Hund's rule coupling between the $3d_{z^2}$ and the $3d_{x^2-y^2}$ orbitals within the same Ni$^{2.5+}$ cation, whose effect will be considered in the present study.

In this Letter, strongly-correlated models are constructed to investigate the pairing mechanism 
to LNO under pressure.
The two half-filled $3d_{z^2}$ orbitals in a unit cell are viewed as two insulating spins which couple via the interlayer AFM superexchange interaction $J_{\perp}$ ~\cite{ZhangGM2023DMRG,Yi_Feng2023}, while the two quarter-filled $3d_{x^2-y^2}$ orbitals take the role of charge carriers.
Under the Hund's rule coupling, the AFM interlayer super-exchange interaction between two $3d_{z^2}$ orbitals 
is transmitted to that between two $3d_{x^2-y^2}$ orbitals
on two Ni sites along a rung.
In combination with the intralayer super-exchange interaction, we arrive at a bilayer $t$-$J$ model \cite{ubbens1994,bohrdt2021exploration,demler2022,hirthe2023magnetically} with the single $3d_{x^2-y^2}$ orbital, which is responsible for the SC in LNO.
Within the slave-boson mean-field (SBMF) theory \cite{kotliar1988,lee2006htsc}, this model is solved to obtain the ground-state phase diagram and superconducting order parameters.
Our result suggests that in the doping regime relevant to experiments, the original intralayer $d$-wave pairing at $J_{\perp}=0$ is changed into the interlayer $s$-wave pairing by a realistic value of $J_{\perp}$.
Adopting realistic parameters obtained from DFT calculations\cite{YaoDX2023}, our results reveal that {the pairing strength} is dramatically enhanced by the interlayer AFM coupling relative to that for the single-layered case, which may well explain the origin of the high $T_c$ SC observed in LNO under pressure\cite{Wang2023LNO}. Our results further suggest that electron doping into the material will
significantly enhance
superconductivity
.

\begin{figure}[t!]
\centering
\includegraphics[width=0.48\textwidth]{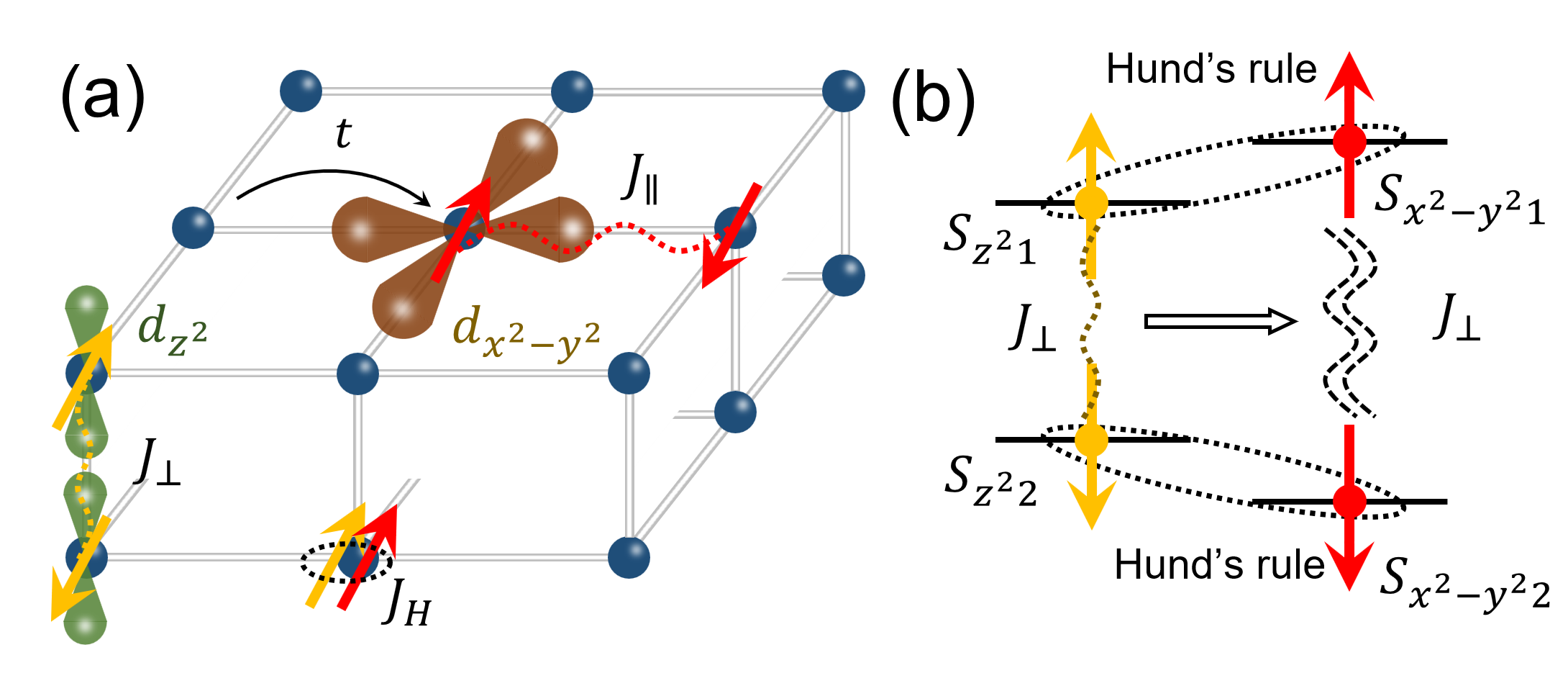}
\caption{(a) Schematic diagram for the two-orbital bilayer  $t$-$J$-$J_H$ model.
The charge carriers reside on the $3d_{x^2-y^2}$ orbitals.
The $3d_{z^2}$ orbital is a singly-occupied localized spin with inter-layer spin exchange $J_{\perp}$.
Two onsite $E_g$ orbitals tend to form a spin-triplet due to the Hund's rule coupling $J_H$. (b) Schematic diagram showing the effective AFM interlayer super-exchange interaction $J_{\perp}$ between the $3d_{x^2-y^2}$ orbitals transmitted from that between the $3d_{z^2}$ orbitals via the on-site Hund's rule coupling $J_H$.
}
\label{fig:LatticeExchange}
\end{figure}

{\bf Model:}
On average the electron numbers in each $3d_{z^2}$ orbital and $3d_{x^2-y^2}$ orbital are $1$ and $0.5$, corresponding to half filling and $1/4$-filling, respectively.
Due to Hund's rule, electrons in $3d_{z^2}$ and $3d_{x^2-y^2}$ orbitals on the same Ni site tend to form a spin-triplet state.
The $3d_{x^2-y^2}$ orbital lies within the NiO$_2$ layer and its interlayer hopping $t_{\perp}$ nearly vanishes.
The bilayer coupling is through the electron hopping of $3d_{z^2}$ orbital, inter-mediated by the $2p$ orbital of interlayer O-atom.
The hopping strength could be significantly enhanced under pressure \cite{Wang2023LNO}.
In the strong coupling limit, the superexchange mechanism induces an effective interlayer AFM spin-exchange $J_{\perp}$ between two $3d_{z^2}$ electrons ~\cite{ZhangGM2023DMRG,Yi_Feng2023}.
The electronic properties 
are described by a two $E_g$-orbitals bilayer $t$-$J$-$J_H$ model, as depicted in Fig.~\ref{fig:LatticeExchange} ($a$).

The model Hamiltonian is $H=H_{\parallel} +H_{\perp}$, with
\begin{align}\label{two_orbital_Hamitonian}
H_{\parallel}
=&-t \sum_{\langle \bm{i},\bm{j} \rangle,\alpha,\sigma}
\big( {c}^{\dagger}_{\bm{i}\alpha\sigma}{c}_{\bm{j}\alpha\sigma}+\text{h.c.} \big)
+J_{\parallel} \sum_{\langle \bm{i},\bm{j} \rangle,\alpha} \bm{S}_{\bm{i}\alpha}\cdot\bm{S}_{\bm{j}\alpha}    \notag\\
 H_{\perp}=&-J_H \sum_{\bm{i},\alpha}
\bm{S}_{\bm{i}\alpha}\cdot \bm{S}_{z^2\bm{i}\alpha}+
J_{\perp} \sum_{\bm{i}} \bm{S}_{z^2\bm{i}1}\cdot \bm{S}_{z^2\bm{i}2}.
\end{align}
Here $c_{\bm{i}\alpha\sigma}^{\dagger}$ creates a $3d_{x^2-y^2}$ electron with spin $\sigma$ on lattice site $\bm{i}$ in the layer $\alpha=1,2$. $\bm{S}_{\bm{i}\alpha}=\frac{1}{2}c_{\bm{i}\alpha\sigma}^{\dagger} [\bm{\sigma}]_{\sigma\sigma^{\prime}}c_{\bm{i}\alpha\sigma^{\prime}}$ is the spin operator for the 3$d_{x^2-y^2}$ orbital, with Pauli matrix $\bm{\sigma}=(\sigma_x,\sigma_y,\sigma_z)$.
The summation $\langle \bm{i}\bm{j}\rangle$ takes over all the nearest-neighboring (NN) bonds.
Hence, $H_{\parallel}$ describes two separate layers of the conventional $t$-$J$ model of $3d_{x^2-y^2}$ electrons with a hopping $t$ term and an AFM spin-exchange $J_{\parallel}$ term.
$\bm{S}_{z^2\bm{i}\alpha}$ is the spin operator of the localized single-occupied $3d_{z^2}$ orbital.
Therefore, $H_{\perp}$ describes the coupling of two $t$-$J$ layers through the Hund's rule coupling $J_H$ between two $E_g$-orbitals
and the interlayer AFM super-exchange $J_{\perp}$ between the $3d_{z^2}$-orbital spins within two layers.

This two-orbital problem can be further simplified into a single $3d_{x^2-y^2}$-orbital one 
since realistically $J_H\gg J_{\parallel,\perp}$.
In this limit, the electron spin in $3d_{x^2-y^2}$
will always be aligned in the directions of $\bm{S}_{z^2\bm{i}\alpha}$ in the same cation, and consequently the AFM interlayer superexchange interaction between the $3d_{z^2}$ spins will be transmitted to  the $3d_{x^2-y^2}$ electrons, as depicted in Fig.~\ref{fig:LatticeExchange} ($b$).
This insight can be verified under the framework of the spin-coherent-state path integral\cite{auerbach1998} given in the Supplemental Material (SM), Sec. A \cite{supp}:
Integrating out the spin degree of freedom of the $3d_{z^2}$ orbital $\bm{S}_{z^2\bm{i}\alpha}$, an effective interlayer spin-exchange between $3d_{x^2-y^2}$ electrons emerges in the semi-classical approximation.
Alternatively, this viewpoint can also be checked in the operator formulation provided in SM, Sec. B \cite{supp}: In the large $J_H$ limit, the $3d_{x^2-y^2}$ and $3d_{z^2}$ orbitals on the same Ni$^{2.5+}$ cation form a spin-triplet.
When acting on this restricted spin-triplet Hilbert space, the spin exchange interaction $\bm{S}_{z^21}\cdot\bm{S}_{z^22}$ is equivalent to $\bm{S}_{1}\cdot\bm{S}_{2}$. The remaining theory is a bilayer single $3d_{x^2-y^2}$-orbital $t$-$J$ model with the nearest-neighbor spin exchange,
\begin{align}\label{Hamiltonian_tJ}
&H
=-t \sum_{\langle \bm{i},\bm{j} \rangle,\alpha,\sigma}
\big( {c}^{\dagger}_{\bm{i}\alpha\sigma}{c}_{\bm{j}\alpha\sigma}+\text{h.c.} \big)
+J_{\parallel} \sum_{\langle \bm{i},\bm{j} \rangle,\alpha} \bm{S}_{\bm{i}\alpha}\cdot\bm{S}_{\bm{j}\alpha}
    \notag\\
&-t_{\perp} \sum_{\bm{i}\sigma}
\big( {c}^{\dagger}_{\bm{i}1\sigma}{c}_{\bm{i}2\sigma}+\text{h.c.} \big)
+J_{\perp} \sum_{\bm{i}} \bm{S}_{\bm{i}1}\cdot \bm{S}_{\bm{i}2}.
\end{align}
Eq.~(\ref{Hamiltonian_tJ}) constitutes a minimal model for the mechanism to SC in LNO.

In the following study, $t$ is set as the energy unit.
The super-exchange interaction $J_{\parallel}=4t^2/U$, where $U$ is the Hubbard interaction
in the $3d_{x^2-y^2}$ orbital.
In this study, $U=10t$ is taken, and hence $J_{\parallel}=0.4t$.
Other choices of $J_{\parallel}$ will not change the conclusions.
As for $J_{\perp}$, we will first set $J_{\perp}/J_{\parallel}$ as a tuning parameter to fully investigate its effect, and then estimate its realistic value from DFT calculations.
In LNO, it is expected that $J_{\perp}/J_{\parallel}>1$. Note that a small inter-layer hopping $t_{\perp}=0.05t$ is added to pin down the relative pairing phase between the two layers.

{\bf The ground-state phase diagram:}
We employ the SBMF theory\cite{kotliar1988,lee2006htsc} to solve with the above bilayer $t$-$J$ model (\ref{Hamiltonian_tJ}) (see SM, Sec.~C \cite{supp} for details).
The electron operator is decomposed into $c_{i\alpha\sigma}^{\dagger}=f_{i\alpha\sigma}^{\dagger}b_{i\alpha}$,
where $f_{i\alpha\sigma}^{\dagger}/b_{i\alpha}$ is the creation/annihilation operator of spinon/holon.
At the mean-field (MF) level, the spinon and holon degrees of freedom are decoupled.
In the ground state, the holons are Bose-Einstein condensed (BEC), and thus their operators are simplified as $b_{i\alpha}=\sqrt{\delta}$, where the hole-doping level $\delta$ is defined as twice of the deviation from half filling and is related to the filling fraction $x$ via $\delta=1-2x$.
In the ideal case, the filling fraction should be $x=0.25$. However, in realistic materials, considering the hybridization between the $3d_{x^2-y^2}$ and the $3d_{z^2}$ orbitals\cite{Yi_Feng2023}, as well as the fact that some holes can reside on the oxygen-anions\cite{WuWei2023charge}, the filling fraction can be above $0.25$.
In this calculation, we set $x\in(0.25,0.35)$, corresponding to $\delta\in(0.3,0.5)$. Note that for the single-layer $t$-$J$ model, the pairing strength in such a heavily overdoped region is very weak \cite{kotliar1988}.

\begin{figure}[t!]
\centering
\includegraphics[width=0.45\textwidth]{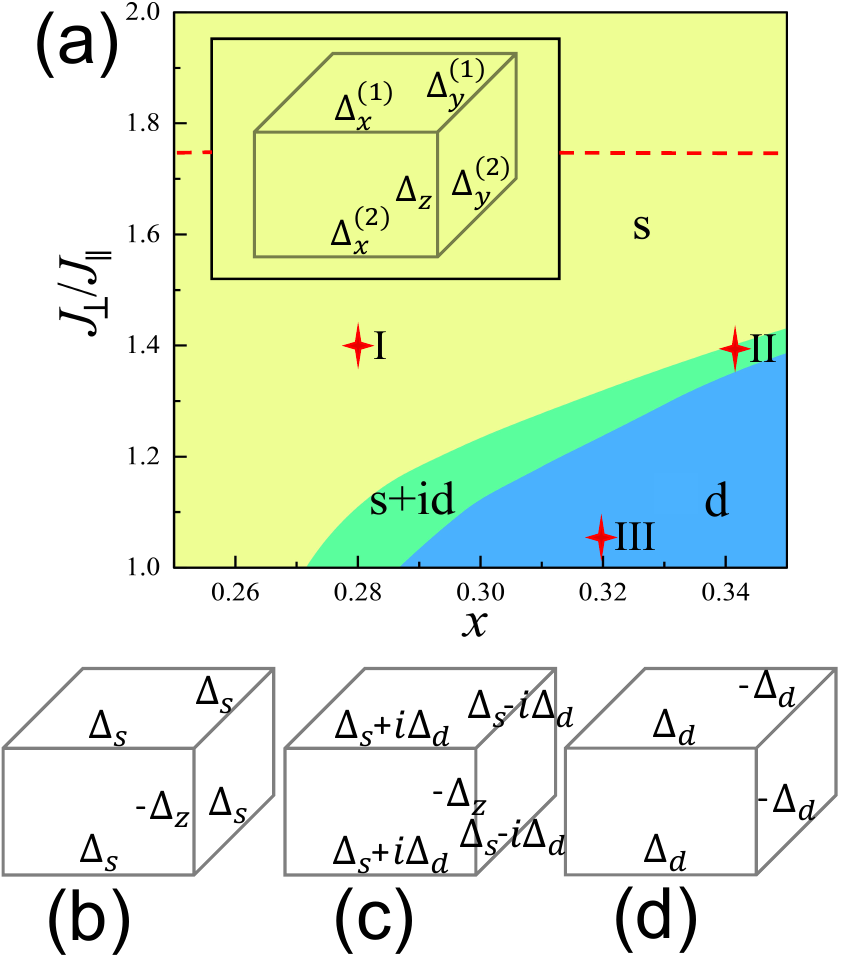}
\caption{(a) Ground state phase diagram with respect to the filling $x$ and $J_{\perp}/J_{\parallel}$ with $J_{\parallel}=0.4t$. The inset shows the pairing order parameters.
At point \textbf{I} (0.28,1.4): $\Delta_z=3.5\times 10^{-3}$, $\Delta^{(1,2)}_x=\Delta^{(1,2)}_y=-2.7\times 10^{-5}$.
At point \textbf{II} (0.342,1.39): $\Delta_z=3.7\times 10^{-2}$, $\Delta^{(1,2)}_x=\Delta^{(1,2)*}_y=(-0.36+4.4i)\times 10^{-3}$.
At point \textbf{III} (0.32,1.05): $\Delta_z$=0, $\Delta^{(1,2)}_x=-\Delta^{(1,2)}_y=5.3\times 10^{-3}$.
The red dashed line marks the realistic $J_{\perp}/J_{\parallel}\approx1.75$ for LNO.
(b-d) show the pairing configurations of the $s$-wave (\textbf{I}), $s+id$-wave (\textbf{II}) and $d$-wave (\textbf{III}), respectively.}
\label{fig:0TPhase}
\end{figure}

The super-exchange terms in Eq. (\ref{Hamiltonian_tJ}) can be decomposed by the following intra- and inter- layer bonding and pairing order parameters,
\begin{equation}
\begin{aligned}
\chi_{\bm{i}\bm{j}}^{(\alpha)}
=&\langle f_{\bm{j}\alpha\uparrow}^{\dagger} f_{\bm{i}\alpha\uparrow}
+f_{\bm{j}\alpha\downarrow}^{\dagger} f_{\bm{i}\alpha\downarrow}\rangle\equiv \chi_{\bm{j}-\bm{i}}^{(\alpha)},   \\
\Delta_{\bm{i}\bm{j}}^{(\alpha)}
=&\langle f_{\bm{j}\alpha\downarrow} f_{\bm{i}\alpha\uparrow}
-f_{\bm{j}\alpha\uparrow} f_{\bm{i}\alpha\downarrow} \rangle
\equiv\Delta_{\bm{j}-\bm{i}}^{(\alpha)},   \\
\chi_{\bm{i},\perp}
=&\langle f_{\bm{i}2\uparrow}^{\dagger} f_{\bm{i}1\uparrow}
+f_{\bm{i}2\downarrow}^{\dagger} f_{\bm{i}1\downarrow}\rangle
\equiv \chi_{z}, \\
\Delta_{\bm{i},\perp}
=&\langle f_{\bm{i}2\downarrow} f_{\bm{i}1\uparrow}
-f_{\bm{i}2\uparrow} f_{\bm{i}1\downarrow} \rangle
\equiv \Delta_{z},
\label{eq:OrderPara}
\end{aligned}
\end{equation}
which are assumed to be site-independent.
The five pairing order parameters are marked in the inset of Fig.~{\ref{fig:0TPhase}}.

The ground-state phase diagram with respect to the filling $x$ and $J_\perp/J_\parallel$ is shown in Fig.~\ref{fig:0TPhase} ($a$).
 As $J_\perp$ should be larger than $J_\parallel$, we have set $J_\perp/J_\parallel\in(1,2)$ in the phase diagram. Three different phases exist in Fig.~\ref{fig:0TPhase} ($a$).
The lower right region (defined as region III) wherein the filling is relatively high and $J_\perp/J_\parallel$ is relatively small is occupied by the $d$-wave pairing.
This region can be continuously connected to the low hole-doped
single-layered $t$-$J$ model representing  cuprates.
The upper left region wherein the filling is relatively low and $J_\perp/J_\parallel$ is relatively large (defined as region I)
shows the $s$-wave pairing.
This region is relevant to LNO, wherein $J_{\perp}/J_{\parallel}\approx1.75$ (red dashed line), see the estimation below.
Note that a variant of the bilayer Hubbard model has been simulated by the sign-free quantum Monte-Carlo approach, also showing the extended $s$-wave pairing \cite{MaTX2022}.
Remarkably, a narrow region (defined as region II) sitting in between region I and III is occupied by the $s+id$-wave pairing.

To gain more information of the pairing nature, one typical point is taken within each region in Fig.~\ref{fig:0TPhase} ($a$) to provide the pairing configurations.
At the typical point in region I showing the $s$-wave pairing, $\Delta_z=3.5\times 10^{-3}$, $\Delta^{(1,2)}_x=\Delta^{(1,2)}_y=-2.7\times 10^{-5}$, schematically shown in Fig.~\ref{fig:0TPhase}($b$).
Consequently, the order parameters in the two layers are in phase, and the interlayer pairing dominates the intralayer one.
It is interesting to note that $\Delta_z$ and $\Delta^{(1,2)}_{x,y}$ hold different signs, which can be thought as the residue of the $d$-wave pairing from the side view.
At the typical point in region III exhibiting the $d$-wave pairing, $\Delta_z$=0, $\Delta^{(1,2)}_x=-\Delta^{(1,2)}_y=5.3\times 10^{-3}$, schematically shown in Fig.~\ref{fig:0TPhase}($d$). It turns out that the $d$-wave pairing order parameters on the two layers are in phase, and the interlayer pairing vanishes as it is inconsistent with this symmetry. At the typical point in region II exhibiting the $s+id$-wave pairing, $\Delta_z=3.7\times 10^{-2}$, $\Delta^{(1,2)}_x=\Delta^{(1,2)*}_y=(-0.36+4.4i)\times 10^{-3}$.
This pairing configuration is schematically shown in Fig.~\ref{fig:0TPhase} ($c$), which can be decomposed as $\Delta_s+i\Delta_d$, wherein the schematic pairing configurations for $\Delta_s$ and  $\Delta_d$ are the same as Fig.~\ref{fig:0TPhase} ($b$) and ($d$).
The $s+id$ pairing state in the intermediate regime spontaneously breaks time-reversal symmetry.
Similar $s+id$ state has been suggested in the much larger filling (or much lower doping) and much smaller $J_{\perp}$ regime\cite{suzumura1988rvb,kuboki1995,zhao2005}.
Such a state could induce non-trivial supercurrent due to spatial inhomogeneity, which can be experimentally detected \cite{LeeWC2009}.

{\bf Interlayer coupling driven SC}:
In the SBMF theory, the onset of SC requires the condensation of holons and the pairing
of spinons.
The ground state SC order parameter is denoted by $\tilde{\Delta}_{\text{SC}}=\delta \Delta_{\text{pair}}$, where $\delta$ represents the holon density and $\Delta_{\text{pair}}$ represents the spinon pairing defined in Eq.~(\ref{eq:OrderPara}).
We would focus on the dominant channel, either the inter- or the intra-layer one.
The obtained dominant $\tilde{\Delta}_{\text{SC}}$ as a function of the filling level $x$ is plotted in Fig.~\ref{fig:JperpTcFilling} ($a$) for various interlayer super-exchange strengths $J_{\perp}/J_{\parallel}$
in comparison to the case of $J_{\perp}=0$.
Obviously, $\tilde{\Delta}_{\text{SC}}$ rises promptly with the increase of $x$ for
all values of $J_{\perp}/J_{\parallel}$.
This feature is similar to the case of $J_\perp=0$ representing the single-layer $t$-$J$ model, wherein $\tilde{\Delta}_{\text{SC}}$
drops rapidly as $\delta$ approaching 0.5, or,
equivalently, $x \to 0.25$.
$\tilde{\Delta}_{\text{SC}}$ as a function of $J_{\perp}/J_{\parallel}$ is shown in Fig.~\ref{fig:JperpTcFilling} ($b$) for specific fillings $x$. 
Notably, $\tilde{\Delta}_{\text{SC}}$ increases monotonically and significantly with the increase of $J_{\perp}/J_{\parallel}$ for $J_{\perp}>J_{\parallel}$ across all these experimentally relevant fillings.

The results shown in Fig.~\ref{fig:JperpTcFilling} ($a$-$b$) are consistent with the important experiment fact that the robust 
SC in LNO only emerges under pressure\cite{Wang2023LNO} because of the following two reasons.
Firstly, pressure enhances the interlayer coupling and hence $J_{\perp}/J_{\parallel}$, leading to the drastic enhancement of $\tilde{\Delta}_{\text{SC}}$ as shown in Fig.~\ref{fig:JperpTcFilling}.
($b$).
Secondly, pressure enhances the hybridization between the $3d_{x^2-y^2}$ and the $3d_{z^2}$ orbitals\cite{Yi_Feng2023}, leading to the increasing of $x$, which also strongly enhances the $\tilde{\Delta}_{\text{SC}}$ as shown in Fig.~\ref{fig:JperpTcFilling} ($a$).
Furthermore, these results are also consistent with the experiment result that the apical-oxygen vacancies suppress SC promptly\cite{Wang2023LNO}.
In our theory, this is because  vacancies break the Ni-O-Ni bonding along the $z$-axis, and hence $J_{\perp}$ vanishes locally, which is harmful for SC.
Besides, Fig.~\ref{fig:JperpTcFilling} ($a$-$b$) further indicate that electron doping into LNO will effectively enhance $\tilde{\Delta}_{\text{SC}}$,
while hole doping will suppress $\tilde{\Delta}_{\text{SC}}$.

The value of $J_{\perp}/J_{\parallel}$ in LNO under pressure can be estimated from DFT calculations.
The interlayer hopping integral of the $3d_{z^2}$ orbital is about 0.635 eV and the intralayer NN hopping integral of the $3d_{x^2-y^2}$ is about 0.48 eV\cite{YaoDX2023}, then $J_{\perp}/J_{\parallel}
\approx 1.75$,
as the Hubbard $U$ of the two orbitals are close.
Fig.~\ref{fig:JperpTcFilling} ($c$) shows the comparison of the filling dependence of $\tilde{\Delta}_{\text{SC}}$ between $J_{\perp}=0$ and the realistic $J_{\perp}$,
which suggests that near $x=0.25$, the $\tilde{\Delta}_{\text{SC}}$ at $J_{\perp}/J_{\parallel}=1.75$ is more than an order of magnitude higher than that at $J_{\perp}=0$.
The pairing symmetry for $J_{\perp}/J_{\parallel}=1.75$ in experimentally relevant filling regime is $s$-wave, consistent with Fig.~\ref{fig:0TPhase} ($a$).
The corresponding pairing configuration is shown in Fig.~\ref{fig:JperpTcFilling} ($d$), wherein $\Delta_x\approx 0$.
Therefore, for these realistic parameters in LNO, the pairing state is the interlayer $s$-wave pairing.

Here the interlayer pairing overwhelms the intralayer one as the former suffers from less pairing frustration than the latter.
For intralayer pairing, an electron has to choose one among the four surrounding NN sites to pair, which compete one another.
Instead for interlayer pairing,
it can focus on the only one 
along the rung to pair.
This not only makes $\Delta_z\gg \Delta_x$, but also greatly enhances $\Delta_z$.
Therefore, the interlayer pairing mechanism leads to the robust interlayer $s$-wave pairing.

\begin{figure}[t!]
\centering
\includegraphics[width=0.48\textwidth]{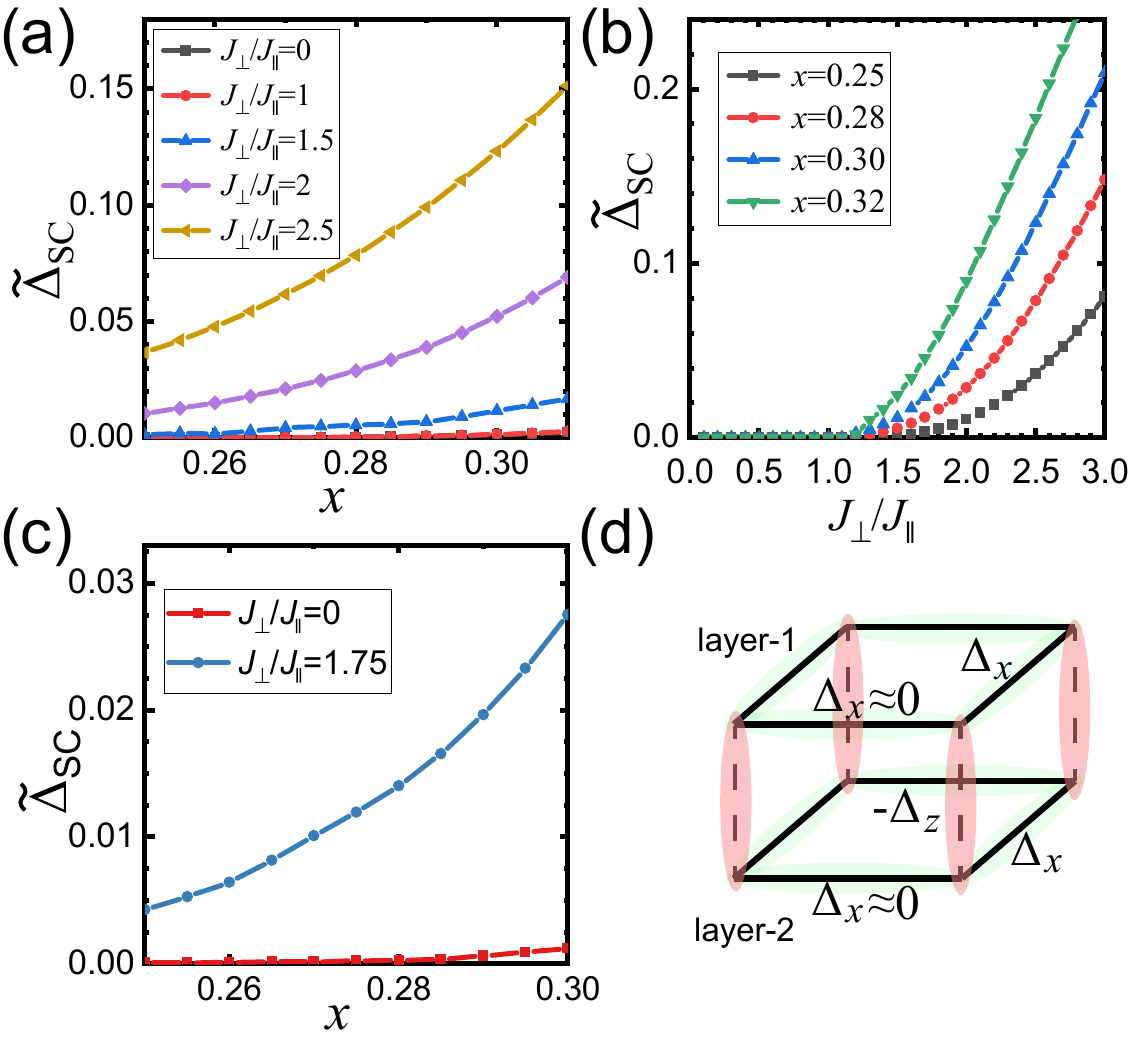}
\caption{(a). Ground state superconducting order parameter $\tilde{\Delta}_{\text{SC}}$ versus filling $x$ at $J_{\parallel}=0.4t$ for different coupling ratios $J_{\perp}/J_{\parallel}=0, 1, 1.5, 2, 2.5$.
(b). $\tilde{\Delta}_{\text{SC}}$ versus $J_{\perp}/J_{\parallel}$ for different filling $x=0.25, 0.28, 0.3, 0.32$.
(c). Comparison of $\tilde{\Delta}_{\text{SC}}$ versus filling $x$ between $J_{\perp}/J_{\parallel}=0$ and $J_{\perp}/J_{\parallel}=1.75$.
(d). Pairing configuration of the obtained interlayer-$s$-wave pairing for $J_{\perp}/J_{\parallel}=1.75$.}
\label{fig:JperpTcFilling}
\end{figure}

In accordance with the strong enhancement of the ground-state $\tilde{\Delta}_{SC}$ by strong $J_{\perp}$,
the superconducting $T_c$ is also strongly enhanced proportionally within the BCS theory.
In the SBMF theory, $T_c$ is given by the lower one of $T_{\text{BEC}}$ and $T_{\text{pair}}$.
Here, $T_{\text{BEC}}$ represents the BEC temperature of holons, which is high due to the large doping level of the $d_{x^2-y^2}$ orbital.
Conversely, $T_{\text{pair}}$, the pairing temperature of spinons, represents the physical $T_c$.
$T_{\text{pair}}$ can be obtained by solving the finite-temperature MF self-consistent gap equation, which turns out to be strongly enhanced by strong $J_{\perp}$ (see SM Sec. D \cite{supp}).
Although the bilayer $t$-$J$ model studied here is a rigorously 2D system, a weak inter-bilayer coupling always exists in real materials, which would stabilize the long-range phase coherence\cite{kopec2000} below a finite $T_c$ near our MF prediction.

{\bf Discussion and conclusion} The role of the ligand oxygen $2p$-orbitals has been so far neglected, while previous studies
proposed that based on the Ni $3d^{8+}$ configuration, i.e., one hole in each Ni $E_g$-orbital, the extra doped holes go to
the ligand $2p$-orbitals
~\cite{mizokawa2000,bisogni2016ligandHole,WuWei2023charge}.
Calculations show that a singly occupied $E_g$-orbital can form a Zhang-Rice singlet (ZRS) with a ligand hole ~\cite{WuWei2023charge}, and
then the ZRS is
mapped to a ``hole" similar to the case of cuprates, otherwise, it is ``occupied".
In this way the Ni-O system is reduced to an effective $E_g$-orbital model with occupations the same as those in Eq.(\ref{two_orbital_Hamitonian}), which justifies our starting point.

In bi- and tri-layer high $T_c$
cuprates ({\it e.g.} YBa$_2$Cu$_3$O$_{7-\delta}$ and Bi$_2$Sr$_2$Ca$_2$Cu$_3$O$_{10+\delta}$),
the $d$-wave pairing takes place inside each layer.
The interlayer Josephson coupling enhances SC phase coherence and increases
$T_c$ by a factor of 2 or 3 \cite{chakravarty2004explanation,chu2015hole,luo2023electronic}.
The situation here is distinct:
The strong interlayer superexchange interaction in the $3d_{x^2-y^2}$ orbital assisted by Hund's rule coupling not only renders the interlayer pairing,
but also strongly enhances the pairing strength and hence the $T_c$.

\section*{Note} Note that a recent study\cite{qu2023bilayer} toward our model Eq. (\ref{Hamiltonian_tJ}) adopting state-of-the-art tensor-network methods has obtained similar results as here.

\section*{Acknowledgments}
We are grateful to Wei Li, Yi-Zhuang You, Yang Qi, Yi-Fan Jiang and Wei-Qiang Chen for stimulating discussions.
F.Y. is supported by the National Natural Science Foundation of China under the Grants No. 12074031, No. 12234016, and No. 11674025.
C.W. is supported by the National Natural Science Foundation of China under the Grants No. 12234016
and No. 12174317. C.L. is supported by the National Natural Science Foundation of China under the Grants No. 12304180.
This work has been supported by the New Cornerstone Science
Foundation.

\twocolumngrid
\bibliography{references}

\begin{thebibliography}{51}%
\makeatletter
\providecommand \@ifxundefined [1]{%
 \@ifx{#1\undefined}
}%
\providecommand \@ifnum [1]{%
 \ifnum #1\expandafter \@firstoftwo
 \else \expandafter \@secondoftwo
 \fi
}%
\providecommand \@ifx [1]{%
 \ifx #1\expandafter \@firstoftwo
 \else \expandafter \@secondoftwo
 \fi
}%
\providecommand \natexlab [1]{#1}%
\providecommand \enquote  [1]{``#1''}%
\providecommand \bibnamefont  [1]{#1}%
\providecommand \bibfnamefont [1]{#1}%
\providecommand \citenamefont [1]{#1}%
\providecommand \href@noop [0]{\@secondoftwo}%
\providecommand \href [0]{\begingroup \@sanitize@url \@href}%
\providecommand \@href[1]{\@@startlink{#1}\@@href}%
\providecommand \@@href[1]{\endgroup#1\@@endlink}%
\providecommand \@sanitize@url [0]{\catcode `\\12\catcode `\$12\catcode
  `\&12\catcode `\#12\catcode `\^12\catcode `\_12\catcode `\%12\relax}%
\providecommand \@@startlink[1]{}%
\providecommand \@@endlink[0]{}%
\providecommand \url  [0]{\begingroup\@sanitize@url \@url }%
\providecommand \@url [1]{\endgroup\@href {#1}{\urlprefix }}%
\providecommand \urlprefix  [0]{URL }%
\providecommand \Eprint [0]{\href }%
\providecommand \doibase [0]{http://dx.doi.org/}%
\providecommand \selectlanguage [0]{\@gobble}%
\providecommand \bibinfo  [0]{\@secondoftwo}%
\providecommand \bibfield  [0]{\@secondoftwo}%
\providecommand \translation [1]{[#1]}%
\providecommand \BibitemOpen [0]{}%
\providecommand \bibitemStop [0]{}%
\providecommand \bibitemNoStop [0]{.\EOS\space}%
\providecommand \EOS [0]{\spacefactor3000\relax}%
\providecommand \BibitemShut  [1]{\csname bibitem#1\endcsname}%
\let\auto@bib@innerbib\@empty
\bibitem [{\citenamefont {Bednorz}\ and\ \citenamefont
  {M{\"u}ller}(1986)}]{bednorz1986LBCO}%
  \BibitemOpen
  \bibfield  {author} {\bibinfo {author} {\bibfnamefont {J.~G.}\ \bibnamefont
  {Bednorz}}\ and\ \bibinfo {author} {\bibfnamefont {K.~A.}\ \bibnamefont
  {M{\"u}ller}},\ }\href {\doibase 10.1007/BF01303701} {\bibfield  {journal}
  {\bibinfo  {journal} {Zeitschrift f{\"u}r Physik B Condensed Matter}\
  }\textbf {\bibinfo {volume} {64}},\ \bibinfo {pages} {189} (\bibinfo {year}
  {1986})}\BibitemShut {NoStop}%
\bibitem [{\citenamefont {Anderson}(1987)}]{anderson1987rvb}%
  \BibitemOpen
  \bibfield  {author} {\bibinfo {author} {\bibfnamefont {P.~W.}\ \bibnamefont
  {Anderson}},\ }\href {\doibase 10.1126/science.235.4793.1196} {\bibfield
  {journal} {\bibinfo  {journal} {science}\ }\textbf {\bibinfo {volume}
  {235}},\ \bibinfo {pages} {1196} (\bibinfo {year} {1987})}\BibitemShut
  {NoStop}%
\bibitem [{\citenamefont {Kotliar}\ and\ \citenamefont
  {Liu}(1988)}]{kotliar1988}%
  \BibitemOpen
  \bibfield  {author} {\bibinfo {author} {\bibfnamefont {G.}~\bibnamefont
  {Kotliar}}\ and\ \bibinfo {author} {\bibfnamefont {J.}~\bibnamefont {Liu}},\
  }\href {\doibase 10.1103/PhysRevB.38.5142} {\bibfield  {journal} {\bibinfo
  {journal} {Phys. Rev. B}\ }\textbf {\bibinfo {volume} {38}},\ \bibinfo
  {pages} {5142} (\bibinfo {year} {1988})}\BibitemShut {NoStop}%
\bibitem [{\citenamefont {Lee}\ \emph {et~al.}(2006)\citenamefont {Lee},
  \citenamefont {Nagaosa},\ and\ \citenamefont {Wen}}]{lee2006htsc}%
  \BibitemOpen
  \bibfield  {author} {\bibinfo {author} {\bibfnamefont {P.~A.}\ \bibnamefont
  {Lee}}, \bibinfo {author} {\bibfnamefont {N.}~\bibnamefont {Nagaosa}}, \ and\
  \bibinfo {author} {\bibfnamefont {X.-G.}\ \bibnamefont {Wen}},\ }\href
  {\doibase 10.1103/RevModPhys.78.17} {\bibfield  {journal} {\bibinfo
  {journal} {Rev. Mod. Phys.}\ }\textbf {\bibinfo {volume} {78}},\ \bibinfo
  {pages} {17} (\bibinfo {year} {2006})}\BibitemShut {NoStop}%
\bibitem [{\citenamefont {Keimer}\ \emph {et~al.}(2015)\citenamefont {Keimer},
  \citenamefont {Kivelson}, \citenamefont {Norman}, \citenamefont {Uchida},\
  and\ \citenamefont {Zaanen}}]{keimer2015highTc}%
  \BibitemOpen
  \bibfield  {author} {\bibinfo {author} {\bibfnamefont {B.}~\bibnamefont
  {Keimer}}, \bibinfo {author} {\bibfnamefont {S.~A.}\ \bibnamefont
  {Kivelson}}, \bibinfo {author} {\bibfnamefont {M.~R.}\ \bibnamefont
  {Norman}}, \bibinfo {author} {\bibfnamefont {S.}~\bibnamefont {Uchida}}, \
  and\ \bibinfo {author} {\bibfnamefont {J.}~\bibnamefont {Zaanen}},\ }\href
  {\doibase 10.1038/nature14165} {\bibfield  {journal} {\bibinfo  {journal}
  {Nature}\ }\textbf {\bibinfo {volume} {518}},\ \bibinfo {pages} {179}
  (\bibinfo {year} {2015})}\BibitemShut {NoStop}%
\bibitem [{\citenamefont {Proust}\ and\ \citenamefont
  {Taillefer}(2019)}]{proust2019highTc}%
  \BibitemOpen
  \bibfield  {author} {\bibinfo {author} {\bibfnamefont {C.}~\bibnamefont
  {Proust}}\ and\ \bibinfo {author} {\bibfnamefont {L.}~\bibnamefont
  {Taillefer}},\ }\href {\doibase 10.1146/annurev-conmatphys-031218-013210}
  {\bibfield  {journal} {\bibinfo  {journal} {Annual Review of Condensed Matter
  Physics}\ }\textbf {\bibinfo {volume} {10}},\ \bibinfo {pages} {409}
  (\bibinfo {year} {2019})}\BibitemShut {NoStop}%
\bibitem [{\citenamefont {Anisimov}\ \emph {et~al.}(1999)\citenamefont
  {Anisimov}, \citenamefont {Bukhvalov},\ and\ \citenamefont
  {Rice}}]{anisimov1999nickelate}%
  \BibitemOpen
  \bibfield  {author} {\bibinfo {author} {\bibfnamefont {V.~I.}\ \bibnamefont
  {Anisimov}}, \bibinfo {author} {\bibfnamefont {D.}~\bibnamefont {Bukhvalov}},
  \ and\ \bibinfo {author} {\bibfnamefont {T.~M.}\ \bibnamefont {Rice}},\
  }\href {\doibase 10.1103/PhysRevB.59.7901} {\bibfield  {journal} {\bibinfo
  {journal} {Phys. Rev. B}\ }\textbf {\bibinfo {volume} {59}},\ \bibinfo
  {pages} {7901} (\bibinfo {year} {1999})}\BibitemShut {NoStop}%
\bibitem [{\citenamefont {Li}\ \emph {et~al.}(2019)\citenamefont {Li},
  \citenamefont {Lee}, \citenamefont {Wang}, \citenamefont {Osada},
  \citenamefont {Crossley}, \citenamefont {Lee}, \citenamefont {Cui},
  \citenamefont {Hikita},\ and\ \citenamefont {Hwang}}]{li2019nickelate}%
  \BibitemOpen
  \bibfield  {author} {\bibinfo {author} {\bibfnamefont {D.}~\bibnamefont
  {Li}}, \bibinfo {author} {\bibfnamefont {K.}~\bibnamefont {Lee}}, \bibinfo
  {author} {\bibfnamefont {B.~Y.}\ \bibnamefont {Wang}}, \bibinfo {author}
  {\bibfnamefont {M.}~\bibnamefont {Osada}}, \bibinfo {author} {\bibfnamefont
  {S.}~\bibnamefont {Crossley}}, \bibinfo {author} {\bibfnamefont {H.~R.}\
  \bibnamefont {Lee}}, \bibinfo {author} {\bibfnamefont {Y.}~\bibnamefont
  {Cui}}, \bibinfo {author} {\bibfnamefont {Y.}~\bibnamefont {Hikita}}, \ and\
  \bibinfo {author} {\bibfnamefont {H.~Y.}\ \bibnamefont {Hwang}},\ }\href
  {\doibase 10.1038/s41586-019-1496-5} {\bibfield  {journal} {\bibinfo
  {journal} {Nature}\ }\textbf {\bibinfo {volume} {572}},\ \bibinfo {pages}
  {624} (\bibinfo {year} {2019})}\BibitemShut {NoStop}%
\bibitem [{\citenamefont {Hu}\ and\ \citenamefont {Wu}(2019)}]{HuLH2019}%
  \BibitemOpen
  \bibfield  {author} {\bibinfo {author} {\bibfnamefont {L.-H.}\ \bibnamefont
  {Hu}}\ and\ \bibinfo {author} {\bibfnamefont {C.}~\bibnamefont {Wu}},\ }\href
  {\doibase 10.1103/PhysRevResearch.1.032046} {\bibfield  {journal} {\bibinfo
  {journal} {Phys. Rev. Res.}\ }\textbf {\bibinfo {volume} {1}},\ \bibinfo
  {pages} {032046} (\bibinfo {year} {2019})}\BibitemShut {NoStop}%
\bibitem [{\citenamefont {Zhang}\ \emph {et~al.}(2020)\citenamefont {Zhang},
  \citenamefont {Yang},\ and\ \citenamefont {Zhang}}]{zhang2020ni}%
  \BibitemOpen
  \bibfield  {author} {\bibinfo {author} {\bibfnamefont {G.-M.}\ \bibnamefont
  {Zhang}}, \bibinfo {author} {\bibfnamefont {Y.-F.}\ \bibnamefont {Yang}}, \
  and\ \bibinfo {author} {\bibfnamefont {F.-C.}\ \bibnamefont {Zhang}},\ }\href
  {\doibase 10.1103/PhysRevB.101.020501} {\bibfield  {journal} {\bibinfo
  {journal} {Phys. Rev. B}\ }\textbf {\bibinfo {volume} {101}},\ \bibinfo
  {pages} {020501} (\bibinfo {year} {2020})}\BibitemShut {NoStop}%
\bibitem [{\citenamefont {Botana}\ \emph {et~al.}(2021)\citenamefont {Botana},
  \citenamefont {Bernardini},\ and\ \citenamefont
  {Cano}}]{botana2021nickelate}%
  \BibitemOpen
  \bibfield  {author} {\bibinfo {author} {\bibfnamefont {A.~S.}\ \bibnamefont
  {Botana}}, \bibinfo {author} {\bibfnamefont {F.}~\bibnamefont {Bernardini}},
  \ and\ \bibinfo {author} {\bibfnamefont {A.}~\bibnamefont {Cano}},\ }\href
  {\doibase 10.1134/S1063776121040026} {\bibfield  {journal} {\bibinfo
  {journal} {Journal of Experimental and Theoretical Physics}\ }\textbf
  {\bibinfo {volume} {132}},\ \bibinfo {pages} {618} (\bibinfo {year}
  {2021})}\BibitemShut {NoStop}%
\bibitem [{\citenamefont {Zeng}\ \emph {et~al.}(2022)\citenamefont {Zeng},
  \citenamefont {Li}, \citenamefont {Chow}, \citenamefont {Cao}, \citenamefont
  {Zhang}, \citenamefont {Tang}, \citenamefont {Yin}, \citenamefont {Lim},
  \citenamefont {Hu}, \citenamefont {Yang},\ and\ \citenamefont
  {Ariando}}]{zeng2022nickelate}%
  \BibitemOpen
  \bibfield  {author} {\bibinfo {author} {\bibfnamefont {S.}~\bibnamefont
  {Zeng}}, \bibinfo {author} {\bibfnamefont {C.}~\bibnamefont {Li}}, \bibinfo
  {author} {\bibfnamefont {L.~E.}\ \bibnamefont {Chow}}, \bibinfo {author}
  {\bibfnamefont {Y.}~\bibnamefont {Cao}}, \bibinfo {author} {\bibfnamefont
  {Z.}~\bibnamefont {Zhang}}, \bibinfo {author} {\bibfnamefont {C.~S.}\
  \bibnamefont {Tang}}, \bibinfo {author} {\bibfnamefont {X.}~\bibnamefont
  {Yin}}, \bibinfo {author} {\bibfnamefont {Z.~S.}\ \bibnamefont {Lim}},
  \bibinfo {author} {\bibfnamefont {J.}~\bibnamefont {Hu}}, \bibinfo {author}
  {\bibfnamefont {P.}~\bibnamefont {Yang}}, \ and\ \bibinfo {author}
  {\bibfnamefont {A.}~\bibnamefont {Ariando}},\ }\href {\doibase
  10.1126/sciadv.abl9927} {\bibfield  {journal} {\bibinfo  {journal} {Science
  advances}\ }\textbf {\bibinfo {volume} {8}},\ \bibinfo {pages} {eabl9927}
  (\bibinfo {year} {2022})}\BibitemShut {NoStop}%
\bibitem [{\citenamefont {Lu}\ \emph {et~al.}(2022)\citenamefont {Lu},
  \citenamefont {Hu}, \citenamefont {Wang}, \citenamefont {Yang},\ and\
  \citenamefont {Wu}}]{LuC2022}%
  \BibitemOpen
  \bibfield  {author} {\bibinfo {author} {\bibfnamefont {C.}~\bibnamefont
  {Lu}}, \bibinfo {author} {\bibfnamefont {L.-H.}\ \bibnamefont {Hu}}, \bibinfo
  {author} {\bibfnamefont {Y.}~\bibnamefont {Wang}}, \bibinfo {author}
  {\bibfnamefont {F.}~\bibnamefont {Yang}}, \ and\ \bibinfo {author}
  {\bibfnamefont {C.}~\bibnamefont {Wu}},\ }\href {\doibase
  10.1103/PhysRevB.105.054516} {\bibfield  {journal} {\bibinfo  {journal}
  {Phys. Rev. B}\ }\textbf {\bibinfo {volume} {105}},\ \bibinfo {pages}
  {054516} (\bibinfo {year} {2022})}\BibitemShut {NoStop}%
\bibitem [{\citenamefont {Sun}\ \emph {et~al.}(2023)\citenamefont {Sun},
  \citenamefont {Huo}, \citenamefont {Hu}, \citenamefont {Li}, \citenamefont
  {Liu}, \citenamefont {Han}, \citenamefont {Tang}, \citenamefont {Mao},
  \citenamefont {Yang}, \citenamefont {Wang}, \citenamefont {Cheng},
  \citenamefont {Yao}, \citenamefont {Zhang},\ and\ \citenamefont
  {Wang}}]{Wang2023LNO}%
  \BibitemOpen
  \bibfield  {author} {\bibinfo {author} {\bibfnamefont {H.}~\bibnamefont
  {Sun}}, \bibinfo {author} {\bibfnamefont {M.}~\bibnamefont {Huo}}, \bibinfo
  {author} {\bibfnamefont {X.}~\bibnamefont {Hu}}, \bibinfo {author}
  {\bibfnamefont {J.}~\bibnamefont {Li}}, \bibinfo {author} {\bibfnamefont
  {Z.}~\bibnamefont {Liu}}, \bibinfo {author} {\bibfnamefont {Y.}~\bibnamefont
  {Han}}, \bibinfo {author} {\bibfnamefont {L.}~\bibnamefont {Tang}}, \bibinfo
  {author} {\bibfnamefont {Z.}~\bibnamefont {Mao}}, \bibinfo {author}
  {\bibfnamefont {P.}~\bibnamefont {Yang}}, \bibinfo {author} {\bibfnamefont
  {B.}~\bibnamefont {Wang}}, \bibinfo {author} {\bibfnamefont {J.}~\bibnamefont
  {Cheng}}, \bibinfo {author} {\bibfnamefont {D.-X.}\ \bibnamefont {Yao}},
  \bibinfo {author} {\bibfnamefont {G.-M.}\ \bibnamefont {Zhang}}, \ and\
  \bibinfo {author} {\bibfnamefont {M.}~\bibnamefont {Wang}},\ }\href {\doibase
  10.1038/s41586-023-06408-7} {\bibfield  {journal} {\bibinfo  {journal}
  {Nature}\ } (\bibinfo {year} {2023}),\
  10.1038/s41586-023-06408-7}\BibitemShut {NoStop}%
\bibitem [{\citenamefont {Liu}\ \emph {et~al.}(2023{\natexlab{a}})\citenamefont
  {Liu}, \citenamefont {Huo}, \citenamefont {Li}, \citenamefont {Li},
  \citenamefont {Liu}, \citenamefont {Dai}, \citenamefont {Zhou}, \citenamefont
  {Hao}, \citenamefont {Lu}, \citenamefont {Wang},\ and\ \citenamefont
  {Wen}}]{WenHH2023}%
  \BibitemOpen
  \bibfield  {author} {\bibinfo {author} {\bibfnamefont {Z.}~\bibnamefont
  {Liu}}, \bibinfo {author} {\bibfnamefont {M.}~\bibnamefont {Huo}}, \bibinfo
  {author} {\bibfnamefont {J.}~\bibnamefont {Li}}, \bibinfo {author}
  {\bibfnamefont {Q.}~\bibnamefont {Li}}, \bibinfo {author} {\bibfnamefont
  {Y.}~\bibnamefont {Liu}}, \bibinfo {author} {\bibfnamefont {Y.}~\bibnamefont
  {Dai}}, \bibinfo {author} {\bibfnamefont {X.}~\bibnamefont {Zhou}}, \bibinfo
  {author} {\bibfnamefont {J.}~\bibnamefont {Hao}}, \bibinfo {author}
  {\bibfnamefont {Y.}~\bibnamefont {Lu}}, \bibinfo {author} {\bibfnamefont
  {M.}~\bibnamefont {Wang}}, \ and\ \bibinfo {author} {\bibfnamefont {H.-H.}\
  \bibnamefont {Wen}},\ }\href {https://arxiv.org/abs/2307.02950} {\bibfield
  {journal} {\bibinfo  {journal} {arXiv preprint arXiv:2307.02950}\ } (\bibinfo
  {year} {2023}{\natexlab{a}})}\BibitemShut {NoStop}%
\bibitem [{\citenamefont {Hou}\ \emph {et~al.}(2023)\citenamefont {Hou},
  \citenamefont {Yang}, \citenamefont {Liu}, \citenamefont {Li}, \citenamefont
  {Shan}, \citenamefont {Ma}, \citenamefont {Wang}, \citenamefont {Wang},
  \citenamefont {Guo}, \citenamefont {Sun}, \citenamefont {Uwatoko},
  \citenamefont {Wang}, \citenamefont {Zhang}, \citenamefont {Wang},\ and\
  \citenamefont {Cheng}}]{Wang2023LNOb}%
  \BibitemOpen
  \bibfield  {author} {\bibinfo {author} {\bibfnamefont {J.}~\bibnamefont
  {Hou}}, \bibinfo {author} {\bibfnamefont {P.-T.}\ \bibnamefont {Yang}},
  \bibinfo {author} {\bibfnamefont {Z.-Y.}\ \bibnamefont {Liu}}, \bibinfo
  {author} {\bibfnamefont {J.-Y.}\ \bibnamefont {Li}}, \bibinfo {author}
  {\bibfnamefont {P.-F.}\ \bibnamefont {Shan}}, \bibinfo {author}
  {\bibfnamefont {L.}~\bibnamefont {Ma}}, \bibinfo {author} {\bibfnamefont
  {G.}~\bibnamefont {Wang}}, \bibinfo {author} {\bibfnamefont {N.-N.}\
  \bibnamefont {Wang}}, \bibinfo {author} {\bibfnamefont {H.-Z.}\ \bibnamefont
  {Guo}}, \bibinfo {author} {\bibfnamefont {J.-P.}\ \bibnamefont {Sun}},
  \bibinfo {author} {\bibfnamefont {Y.}~\bibnamefont {Uwatoko}}, \bibinfo
  {author} {\bibfnamefont {M.}~\bibnamefont {Wang}}, \bibinfo {author}
  {\bibfnamefont {G.-M.}\ \bibnamefont {Zhang}}, \bibinfo {author}
  {\bibfnamefont {B.-S.}\ \bibnamefont {Wang}}, \ and\ \bibinfo {author}
  {\bibfnamefont {J.-G.}\ \bibnamefont {Cheng}},\ }\href {\doibase
  10.1088/0256-307X/40/11/117302} {\bibfield  {journal} {\bibinfo  {journal}
  {Chinese Physics Letters}\ }\textbf {\bibinfo {volume} {40}},\ \bibinfo {eid}
  {117302} (\bibinfo {year} {2023})}\BibitemShut {NoStop}%
\bibitem [{\citenamefont {Zhang}\ \emph
  {et~al.}(2023{\natexlab{a}})\citenamefont {Zhang}, \citenamefont {Su},
  \citenamefont {Huang}, \citenamefont {Sun}, \citenamefont {Huo},
  \citenamefont {Shan}, \citenamefont {Ye}, \citenamefont {Yang}, \citenamefont
  {Li}, \citenamefont {Smidman}, \citenamefont {Wang}, \citenamefont {Jiao},\
  and\ \citenamefont {Yuan}}]{YuanHQ2023LNO}%
  \BibitemOpen
  \bibfield  {author} {\bibinfo {author} {\bibfnamefont {Y.}~\bibnamefont
  {Zhang}}, \bibinfo {author} {\bibfnamefont {D.}~\bibnamefont {Su}}, \bibinfo
  {author} {\bibfnamefont {Y.}~\bibnamefont {Huang}}, \bibinfo {author}
  {\bibfnamefont {H.}~\bibnamefont {Sun}}, \bibinfo {author} {\bibfnamefont
  {M.}~\bibnamefont {Huo}}, \bibinfo {author} {\bibfnamefont {Z.}~\bibnamefont
  {Shan}}, \bibinfo {author} {\bibfnamefont {K.}~\bibnamefont {Ye}}, \bibinfo
  {author} {\bibfnamefont {Z.}~\bibnamefont {Yang}}, \bibinfo {author}
  {\bibfnamefont {R.}~\bibnamefont {Li}}, \bibinfo {author} {\bibfnamefont
  {M.}~\bibnamefont {Smidman}}, \bibinfo {author} {\bibfnamefont
  {M.}~\bibnamefont {Wang}}, \bibinfo {author} {\bibfnamefont {L.}~\bibnamefont
  {Jiao}}, \ and\ \bibinfo {author} {\bibfnamefont {H.}~\bibnamefont {Yuan}},\
  }\href {https://arxiv.org/abs/2307.14819} {\bibfield  {journal} {\bibinfo
  {journal} {arXiv preprint arXiv:2307.14819}\ } (\bibinfo {year}
  {2023}{\natexlab{a}})}\BibitemShut {NoStop}%
\bibitem [{\citenamefont {Luo}\ \emph {et~al.}(2023{\natexlab{a}})\citenamefont
  {Luo}, \citenamefont {Hu}, \citenamefont {Wang}, \citenamefont {W\'u},\ and\
  \citenamefont {Yao}}]{YaoDX2023}%
  \BibitemOpen
  \bibfield  {author} {\bibinfo {author} {\bibfnamefont {Z.}~\bibnamefont
  {Luo}}, \bibinfo {author} {\bibfnamefont {X.}~\bibnamefont {Hu}}, \bibinfo
  {author} {\bibfnamefont {M.}~\bibnamefont {Wang}}, \bibinfo {author}
  {\bibfnamefont {W.}~\bibnamefont {W\'u}}, \ and\ \bibinfo {author}
  {\bibfnamefont {D.-X.}\ \bibnamefont {Yao}},\ }\href {\doibase
  10.1103/PhysRevLett.131.126001} {\bibfield  {journal} {\bibinfo  {journal}
  {Phys. Rev. Lett.}\ }\textbf {\bibinfo {volume} {131}},\ \bibinfo {pages}
  {126001} (\bibinfo {year} {2023}{\natexlab{a}})}\BibitemShut {NoStop}%
\bibitem [{\citenamefont {Zhang}\ \emph
  {et~al.}(2023{\natexlab{b}})\citenamefont {Zhang}, \citenamefont {Lin},
  \citenamefont {Moreo},\ and\ \citenamefont {Dagotto}}]{Dagotto2023}%
  \BibitemOpen
  \bibfield  {author} {\bibinfo {author} {\bibfnamefont {Y.}~\bibnamefont
  {Zhang}}, \bibinfo {author} {\bibfnamefont {L.-F.}\ \bibnamefont {Lin}},
  \bibinfo {author} {\bibfnamefont {A.}~\bibnamefont {Moreo}}, \ and\ \bibinfo
  {author} {\bibfnamefont {E.}~\bibnamefont {Dagotto}},\ }\href {\doibase
  10.1103/PhysRevB.108.L180510} {\bibfield  {journal} {\bibinfo  {journal}
  {Phys. Rev. B}\ }\textbf {\bibinfo {volume} {108}},\ \bibinfo {pages}
  {L180510} (\bibinfo {year} {2023}{\natexlab{b}})}\BibitemShut {NoStop}%
\bibitem [{\citenamefont {Yang}\ \emph
  {et~al.}(2023{\natexlab{a}})\citenamefont {Yang}, \citenamefont {Wang},\ and\
  \citenamefont {Wang}}]{WangQH2023}%
  \BibitemOpen
  \bibfield  {author} {\bibinfo {author} {\bibfnamefont {Q.-G.}\ \bibnamefont
  {Yang}}, \bibinfo {author} {\bibfnamefont {D.}~\bibnamefont {Wang}}, \ and\
  \bibinfo {author} {\bibfnamefont {Q.-H.}\ \bibnamefont {Wang}},\ }\href
  {\doibase 10.1103/PhysRevB.108.L140505} {\bibfield  {journal} {\bibinfo
  {journal} {Phys. Rev. B}\ }\textbf {\bibinfo {volume} {108}},\ \bibinfo
  {pages} {L140505} (\bibinfo {year} {2023}{\natexlab{a}})}\BibitemShut
  {NoStop}%
\bibitem [{\citenamefont {Lechermann}\ \emph {et~al.}(2023)\citenamefont
  {Lechermann}, \citenamefont {Gondolf}, \citenamefont {B\"otzel},\ and\
  \citenamefont {Eremin}}]{lechermann2023}%
  \BibitemOpen
  \bibfield  {author} {\bibinfo {author} {\bibfnamefont {F.}~\bibnamefont
  {Lechermann}}, \bibinfo {author} {\bibfnamefont {J.}~\bibnamefont {Gondolf}},
  \bibinfo {author} {\bibfnamefont {S.}~\bibnamefont {B\"otzel}}, \ and\
  \bibinfo {author} {\bibfnamefont {I.~M.}\ \bibnamefont {Eremin}},\ }\href
  {\doibase 10.1103/PhysRevB.108.L201121} {\bibfield  {journal} {\bibinfo
  {journal} {Phys. Rev. B}\ }\textbf {\bibinfo {volume} {108}},\ \bibinfo
  {pages} {L201121} (\bibinfo {year} {2023})}\BibitemShut {NoStop}%
\bibitem [{\citenamefont {Sakakibara}\ \emph {et~al.}(2023)\citenamefont
  {Sakakibara}, \citenamefont {Kitamine}, \citenamefont {Ochi},\ and\
  \citenamefont {Kuroki}}]{Kuroki2023}%
  \BibitemOpen
  \bibfield  {author} {\bibinfo {author} {\bibfnamefont {H.}~\bibnamefont
  {Sakakibara}}, \bibinfo {author} {\bibfnamefont {N.}~\bibnamefont
  {Kitamine}}, \bibinfo {author} {\bibfnamefont {M.}~\bibnamefont {Ochi}}, \
  and\ \bibinfo {author} {\bibfnamefont {K.}~\bibnamefont {Kuroki}},\ }\href
  {https://arxiv.org/abs/2306.06039} {\bibfield  {journal} {\bibinfo  {journal}
  {arXiv preprint arXiv:2306.06039}\ } (\bibinfo {year} {2023})}\BibitemShut
  {NoStop}%
\bibitem [{\citenamefont {Gu}\ \emph {et~al.}(2023)\citenamefont {Gu},
  \citenamefont {Le}, \citenamefont {Yang}, \citenamefont {Wu},\ and\
  \citenamefont {Hu}}]{HuJP2023}%
  \BibitemOpen
  \bibfield  {author} {\bibinfo {author} {\bibfnamefont {Y.}~\bibnamefont
  {Gu}}, \bibinfo {author} {\bibfnamefont {C.}~\bibnamefont {Le}}, \bibinfo
  {author} {\bibfnamefont {Z.}~\bibnamefont {Yang}}, \bibinfo {author}
  {\bibfnamefont {X.}~\bibnamefont {Wu}}, \ and\ \bibinfo {author}
  {\bibfnamefont {J.}~\bibnamefont {Hu}},\ }\href
  {https://arxiv.org/abs/2306.07275} {\bibfield  {journal} {\bibinfo  {journal}
  {arXiv preprint arXiv:2306.07275}\ } (\bibinfo {year} {2023})}\BibitemShut
  {NoStop}%
\bibitem [{\citenamefont {Shen}\ \emph {et~al.}(2023)\citenamefont {Shen},
  \citenamefont {Qin},\ and\ \citenamefont {Zhang}}]{ZhangGM2023DMRG}%
  \BibitemOpen
  \bibfield  {author} {\bibinfo {author} {\bibfnamefont {Y.}~\bibnamefont
  {Shen}}, \bibinfo {author} {\bibfnamefont {M.}~\bibnamefont {Qin}}, \ and\
  \bibinfo {author} {\bibfnamefont {G.-M.}\ \bibnamefont {Zhang}},\ }\href
  {https://iopscience.iop.org/article/10.1088/0256-307X/40/12/127401}
  {\bibfield  {journal} {\bibinfo  {journal} {Chinese Physics Letters}\
  }\textbf {\bibinfo {volume} {40}},\ \bibinfo {pages} {127401} (\bibinfo
  {year} {2023})}\BibitemShut {NoStop}%
\bibitem [{\citenamefont {Christiansson}\ \emph {et~al.}(2023)\citenamefont
  {Christiansson}, \citenamefont {Petocchi},\ and\ \citenamefont
  {Werner}}]{Werner2023}%
  \BibitemOpen
  \bibfield  {author} {\bibinfo {author} {\bibfnamefont {V.}~\bibnamefont
  {Christiansson}}, \bibinfo {author} {\bibfnamefont {F.}~\bibnamefont
  {Petocchi}}, \ and\ \bibinfo {author} {\bibfnamefont {P.}~\bibnamefont
  {Werner}},\ }\href {\doibase 10.1103/PhysRevLett.131.206501} {\bibfield
  {journal} {\bibinfo  {journal} {Phys. Rev. Lett.}\ }\textbf {\bibinfo
  {volume} {131}},\ \bibinfo {pages} {206501} (\bibinfo {year}
  {2023})}\BibitemShut {NoStop}%
\bibitem [{\citenamefont {Shilenko}\ and\ \citenamefont
  {Leonov}(2023)}]{shilenko2023correlated}%
  \BibitemOpen
  \bibfield  {author} {\bibinfo {author} {\bibfnamefont {D.~A.}\ \bibnamefont
  {Shilenko}}\ and\ \bibinfo {author} {\bibfnamefont {I.~V.}\ \bibnamefont
  {Leonov}},\ }\href {\doibase 10.1103/PhysRevB.108.125105} {\bibfield
  {journal} {\bibinfo  {journal} {Phys. Rev. B}\ }\textbf {\bibinfo {volume}
  {108}},\ \bibinfo {pages} {125105} (\bibinfo {year} {2023})}\BibitemShut
  {NoStop}%
\bibitem [{\citenamefont {W{\'u}}\ \emph {et~al.}(2023)\citenamefont {W{\'u}},
  \citenamefont {Luo}, \citenamefont {Yao},\ and\ \citenamefont
  {Wang}}]{WuWei2023charge}%
  \BibitemOpen
  \bibfield  {author} {\bibinfo {author} {\bibfnamefont {W.}~\bibnamefont
  {W{\'u}}}, \bibinfo {author} {\bibfnamefont {Z.}~\bibnamefont {Luo}},
  \bibinfo {author} {\bibfnamefont {D.-X.}\ \bibnamefont {Yao}}, \ and\
  \bibinfo {author} {\bibfnamefont {M.}~\bibnamefont {Wang}},\ }\href
  {https://arxiv.org/abs/2307.05662} {\bibfield  {journal} {\bibinfo  {journal}
  {arXiv preprint arXiv:2307.05662}\ } (\bibinfo {year} {2023})}\BibitemShut
  {NoStop}%
\bibitem [{\citenamefont {Cao}\ and\ \citenamefont {Yang}(2024)}]{cao2023flat}%
  \BibitemOpen
  \bibfield  {author} {\bibinfo {author} {\bibfnamefont {Y.}~\bibnamefont
  {Cao}}\ and\ \bibinfo {author} {\bibfnamefont {Y.-f.}\ \bibnamefont {Yang}},\
  }\href {\doibase 10.1103/PhysRevB.109.L081105} {\bibfield  {journal}
  {\bibinfo  {journal} {Phys. Rev. B}\ }\textbf {\bibinfo {volume} {109}},\
  \bibinfo {pages} {L081105} (\bibinfo {year} {2024})}\BibitemShut {NoStop}%
\bibitem [{\citenamefont {Chen}\ \emph {et~al.}(2023)\citenamefont {Chen},
  \citenamefont {Jiang}, \citenamefont {Li}, \citenamefont {Zhong},\ and\
  \citenamefont {Lu}}]{chen2023critical}%
  \BibitemOpen
  \bibfield  {author} {\bibinfo {author} {\bibfnamefont {X.}~\bibnamefont
  {Chen}}, \bibinfo {author} {\bibfnamefont {P.}~\bibnamefont {Jiang}},
  \bibinfo {author} {\bibfnamefont {J.}~\bibnamefont {Li}}, \bibinfo {author}
  {\bibfnamefont {Z.}~\bibnamefont {Zhong}}, \ and\ \bibinfo {author}
  {\bibfnamefont {Y.}~\bibnamefont {Lu}},\ }\href
  {https://arxiv.org/abs/2307.07154} {\bibfield  {journal} {\bibinfo  {journal}
  {arXiv preprint arXiv:2307.07154}\ } (\bibinfo {year} {2023})}\BibitemShut
  {NoStop}%
\bibitem [{\citenamefont {Liu}\ \emph {et~al.}(2023{\natexlab{b}})\citenamefont
  {Liu}, \citenamefont {Mei}, \citenamefont {Ye}, \citenamefont {Chen},\ and\
  \citenamefont {Yang}}]{YangF2023}%
  \BibitemOpen
  \bibfield  {author} {\bibinfo {author} {\bibfnamefont {Y.-B.}\ \bibnamefont
  {Liu}}, \bibinfo {author} {\bibfnamefont {J.-W.}\ \bibnamefont {Mei}},
  \bibinfo {author} {\bibfnamefont {F.}~\bibnamefont {Ye}}, \bibinfo {author}
  {\bibfnamefont {W.-Q.}\ \bibnamefont {Chen}}, \ and\ \bibinfo {author}
  {\bibfnamefont {F.}~\bibnamefont {Yang}},\ }\href {\doibase
  10.1103/PhysRevLett.131.236002} {\bibfield  {journal} {\bibinfo  {journal}
  {Phys. Rev. Lett.}\ }\textbf {\bibinfo {volume} {131}},\ \bibinfo {pages}
  {236002} (\bibinfo {year} {2023}{\natexlab{b}})}\BibitemShut {NoStop}%
\bibitem [{\citenamefont {Oh}\ and\ \citenamefont {Zhang}(2023)}]{oh2023type2}%
  \BibitemOpen
  \bibfield  {author} {\bibinfo {author} {\bibfnamefont {H.}~\bibnamefont
  {Oh}}\ and\ \bibinfo {author} {\bibfnamefont {Y.-H.}\ \bibnamefont {Zhang}},\
  }\href {\doibase 10.1103/PhysRevB.108.174511} {\bibfield  {journal} {\bibinfo
   {journal} {Phys. Rev. B}\ }\textbf {\bibinfo {volume} {108}},\ \bibinfo
  {pages} {174511} (\bibinfo {year} {2023})}\BibitemShut {NoStop}%
\bibitem [{\citenamefont {Qu}\ \emph {et~al.}(2024)\citenamefont {Qu},
  \citenamefont {Qu}, \citenamefont {Chen}, \citenamefont {Wu}, \citenamefont
  {Yang}, \citenamefont {Li},\ and\ \citenamefont {Su}}]{qu2023bilayer}%
  \BibitemOpen
  \bibfield  {author} {\bibinfo {author} {\bibfnamefont {X.-Z.}\ \bibnamefont
  {Qu}}, \bibinfo {author} {\bibfnamefont {D.-W.}\ \bibnamefont {Qu}}, \bibinfo
  {author} {\bibfnamefont {J.}~\bibnamefont {Chen}}, \bibinfo {author}
  {\bibfnamefont {C.}~\bibnamefont {Wu}}, \bibinfo {author} {\bibfnamefont
  {F.}~\bibnamefont {Yang}}, \bibinfo {author} {\bibfnamefont {W.}~\bibnamefont
  {Li}}, \ and\ \bibinfo {author} {\bibfnamefont {G.}~\bibnamefont {Su}},\
  }\href {\doibase 10.1103/PhysRevLett.132.036502} {\bibfield  {journal}
  {\bibinfo  {journal} {Phys. Rev. Lett.}\ }\textbf {\bibinfo {volume} {132}},\
  \bibinfo {pages} {036502} (\bibinfo {year} {2024})}\BibitemShut {NoStop}%
\bibitem [{\citenamefont {Yang}\ \emph
  {et~al.}(2023{\natexlab{b}})\citenamefont {Yang}, \citenamefont {Zhang},\
  and\ \citenamefont {Zhang}}]{Yi_Feng2023}%
  \BibitemOpen
  \bibfield  {author} {\bibinfo {author} {\bibfnamefont {Y.-f.}\ \bibnamefont
  {Yang}}, \bibinfo {author} {\bibfnamefont {G.-M.}\ \bibnamefont {Zhang}}, \
  and\ \bibinfo {author} {\bibfnamefont {F.-C.}\ \bibnamefont {Zhang}},\ }\href
  {\doibase 10.1103/PhysRevB.108.L201108} {\bibfield  {journal} {\bibinfo
  {journal} {Phys. Rev. B}\ }\textbf {\bibinfo {volume} {108}},\ \bibinfo
  {pages} {L201108} (\bibinfo {year} {2023}{\natexlab{b}})}\BibitemShut
  {NoStop}%
\bibitem [{\citenamefont {Pardo}\ and\ \citenamefont
  {Pickett}(2011)}]{pardo2011dft}%
  \BibitemOpen
  \bibfield  {author} {\bibinfo {author} {\bibfnamefont {V.}~\bibnamefont
  {Pardo}}\ and\ \bibinfo {author} {\bibfnamefont {W.~E.}\ \bibnamefont
  {Pickett}},\ }\href {\doibase 10.1103/PhysRevB.83.245128} {\bibfield
  {journal} {\bibinfo  {journal} {Phys. Rev. B}\ }\textbf {\bibinfo {volume}
  {83}},\ \bibinfo {pages} {245128} (\bibinfo {year} {2011})}\BibitemShut
  {NoStop}%
\bibitem [{\citenamefont {Ubbens}\ and\ \citenamefont
  {Lee}(1994)}]{ubbens1994}%
  \BibitemOpen
  \bibfield  {author} {\bibinfo {author} {\bibfnamefont {M.~U.}\ \bibnamefont
  {Ubbens}}\ and\ \bibinfo {author} {\bibfnamefont {P.~A.}\ \bibnamefont
  {Lee}},\ }\href {\doibase 10.1103/PhysRevB.50.438} {\bibfield  {journal}
  {\bibinfo  {journal} {Phys. Rev. B}\ }\textbf {\bibinfo {volume} {50}},\
  \bibinfo {pages} {438} (\bibinfo {year} {1994})}\BibitemShut {NoStop}%
\bibitem [{\citenamefont {Bohrdt}\ \emph {et~al.}(2021)\citenamefont {Bohrdt},
  \citenamefont {Homeier}, \citenamefont {Reinmoser}, \citenamefont {Demler},\
  and\ \citenamefont {Grusdt}}]{bohrdt2021exploration}%
  \BibitemOpen
  \bibfield  {author} {\bibinfo {author} {\bibfnamefont {A.}~\bibnamefont
  {Bohrdt}}, \bibinfo {author} {\bibfnamefont {L.}~\bibnamefont {Homeier}},
  \bibinfo {author} {\bibfnamefont {C.}~\bibnamefont {Reinmoser}}, \bibinfo
  {author} {\bibfnamefont {E.}~\bibnamefont {Demler}}, \ and\ \bibinfo {author}
  {\bibfnamefont {F.}~\bibnamefont {Grusdt}},\ }\href
  {https://www.sciencedirect.com/science/article/pii/S0003491621002578}
  {\bibfield  {journal} {\bibinfo  {journal} {Annals of Physics}\ }\textbf
  {\bibinfo {volume} {435}},\ \bibinfo {pages} {168651} (\bibinfo {year}
  {2021})}\BibitemShut {NoStop}%
\bibitem [{\citenamefont {Bohrdt}\ \emph {et~al.}(2022)\citenamefont {Bohrdt},
  \citenamefont {Homeier}, \citenamefont {Bloch}, \citenamefont {Demler},\ and\
  \citenamefont {Grusdt}}]{demler2022}%
  \BibitemOpen
  \bibfield  {author} {\bibinfo {author} {\bibfnamefont {A.}~\bibnamefont
  {Bohrdt}}, \bibinfo {author} {\bibfnamefont {L.}~\bibnamefont {Homeier}},
  \bibinfo {author} {\bibfnamefont {I.}~\bibnamefont {Bloch}}, \bibinfo
  {author} {\bibfnamefont {E.}~\bibnamefont {Demler}}, \ and\ \bibinfo {author}
  {\bibfnamefont {F.}~\bibnamefont {Grusdt}},\ }\href
  {https://www.nature.com/articles/s41567-022-01561-8} {\bibfield  {journal}
  {\bibinfo  {journal} {Nature Physics}\ }\textbf {\bibinfo {volume} {18}},\
  \bibinfo {pages} {651} (\bibinfo {year} {2022})}\BibitemShut {NoStop}%
\bibitem [{\citenamefont {Hirthe}\ \emph {et~al.}(2023)\citenamefont {Hirthe},
  \citenamefont {Chalopin}, \citenamefont {Bourgund}, \citenamefont
  {Bojovi{\'c}}, \citenamefont {Bohrdt}, \citenamefont {Demler}, \citenamefont
  {Grusdt}, \citenamefont {Bloch},\ and\ \citenamefont
  {Hilker}}]{hirthe2023magnetically}%
  \BibitemOpen
  \bibfield  {author} {\bibinfo {author} {\bibfnamefont {S.}~\bibnamefont
  {Hirthe}}, \bibinfo {author} {\bibfnamefont {T.}~\bibnamefont {Chalopin}},
  \bibinfo {author} {\bibfnamefont {D.}~\bibnamefont {Bourgund}}, \bibinfo
  {author} {\bibfnamefont {P.}~\bibnamefont {Bojovi{\'c}}}, \bibinfo {author}
  {\bibfnamefont {A.}~\bibnamefont {Bohrdt}}, \bibinfo {author} {\bibfnamefont
  {E.}~\bibnamefont {Demler}}, \bibinfo {author} {\bibfnamefont
  {F.}~\bibnamefont {Grusdt}}, \bibinfo {author} {\bibfnamefont
  {I.}~\bibnamefont {Bloch}}, \ and\ \bibinfo {author} {\bibfnamefont {T.~A.}\
  \bibnamefont {Hilker}},\ }\href
  {https://www.nature.com/articles/s41586-022-05437-y} {\bibfield  {journal}
  {\bibinfo  {journal} {Nature}\ }\textbf {\bibinfo {volume} {613}},\ \bibinfo
  {pages} {463} (\bibinfo {year} {2023})}\BibitemShut {NoStop}%
\bibitem [{\citenamefont {Auerbach}(1998)}]{auerbach1998}%
  \BibitemOpen
  \bibfield  {author} {\bibinfo {author} {\bibfnamefont {A.}~\bibnamefont
  {Auerbach}},\ }\href@noop {} {\emph {\bibinfo {title} {Interacting electrons
  and quantum magnetism}}},\ \bibinfo {edition} {corr., 2. print}\ ed.,\
  Graduate texts in contemporary physics\ (\bibinfo  {publisher} {Springer},\
  \bibinfo {address} {New York Berlin Heidelberg},\ \bibinfo {year}
  {1998})\BibitemShut {NoStop}%
\bibitem [{sup()}]{supp}%
  \BibitemOpen
  \href@noop {} {}\bibinfo {note} {See Supplemental Material at (URL will be
  inserted by publisher) for additional information.}\BibitemShut {Stop}%
\bibitem [{\citenamefont {Ma}\ \emph {et~al.}(2022)\citenamefont {Ma},
  \citenamefont {Wang},\ and\ \citenamefont {Wu}}]{MaTX2022}%
  \BibitemOpen
  \bibfield  {author} {\bibinfo {author} {\bibfnamefont {T.}~\bibnamefont
  {Ma}}, \bibinfo {author} {\bibfnamefont {D.}~\bibnamefont {Wang}}, \ and\
  \bibinfo {author} {\bibfnamefont {C.}~\bibnamefont {Wu}},\ }\href {\doibase
  10.1103/PhysRevB.106.054510} {\bibfield  {journal} {\bibinfo  {journal}
  {Phys. Rev. B}\ }\textbf {\bibinfo {volume} {106}},\ \bibinfo {pages}
  {054510} (\bibinfo {year} {2022})}\BibitemShut {NoStop}%
\bibitem [{\citenamefont {Suzumura}\ \emph {et~al.}(1988)\citenamefont
  {Suzumura}, \citenamefont {Hasegawa},\ and\ \citenamefont
  {Fukuyama}}]{suzumura1988rvb}%
  \BibitemOpen
  \bibfield  {author} {\bibinfo {author} {\bibfnamefont {Y.}~\bibnamefont
  {Suzumura}}, \bibinfo {author} {\bibfnamefont {Y.}~\bibnamefont {Hasegawa}},
  \ and\ \bibinfo {author} {\bibfnamefont {H.}~\bibnamefont {Fukuyama}},\
  }\href@noop {} {\bibfield  {journal} {\bibinfo  {journal} {Journal of the
  Physical Society of Japan}\ }\textbf {\bibinfo {volume} {57}},\ \bibinfo
  {pages} {2768} (\bibinfo {year} {1988})}\BibitemShut {NoStop}%
\bibitem [{\citenamefont {Kuboki}\ and\ \citenamefont
  {A.~Lee}(1995)}]{kuboki1995}%
  \BibitemOpen
  \bibfield  {author} {\bibinfo {author} {\bibfnamefont {K.}~\bibnamefont
  {Kuboki}}\ and\ \bibinfo {author} {\bibfnamefont {P.}~\bibnamefont
  {A.~Lee}},\ }\href@noop {} {\bibfield  {journal} {\bibinfo  {journal}
  {Journal of the Physical Society of Japan}\ }\textbf {\bibinfo {volume}
  {64}},\ \bibinfo {pages} {3179} (\bibinfo {year} {1995})}\BibitemShut
  {NoStop}%
\bibitem [{\citenamefont {Zhao}\ and\ \citenamefont
  {Engelbrecht}(2005)}]{zhao2005}%
  \BibitemOpen
  \bibfield  {author} {\bibinfo {author} {\bibfnamefont {H.}~\bibnamefont
  {Zhao}}\ and\ \bibinfo {author} {\bibfnamefont {J.~R.}\ \bibnamefont
  {Engelbrecht}},\ }\href {\doibase 10.1103/PhysRevB.71.054508} {\bibfield
  {journal} {\bibinfo  {journal} {Phys. Rev. B}\ }\textbf {\bibinfo {volume}
  {71}},\ \bibinfo {pages} {054508} (\bibinfo {year} {2005})}\BibitemShut
  {NoStop}%
\bibitem [{\citenamefont {Lee}\ \emph {et~al.}(2009)\citenamefont {Lee},
  \citenamefont {Zhang},\ and\ \citenamefont {Wu}}]{LeeWC2009}%
  \BibitemOpen
  \bibfield  {author} {\bibinfo {author} {\bibfnamefont {W.-C.}\ \bibnamefont
  {Lee}}, \bibinfo {author} {\bibfnamefont {S.-C.}\ \bibnamefont {Zhang}}, \
  and\ \bibinfo {author} {\bibfnamefont {C.}~\bibnamefont {Wu}},\ }\href
  {\doibase 10.1103/PhysRevLett.102.217002} {\bibfield  {journal} {\bibinfo
  {journal} {Phys. Rev. Lett.}\ }\textbf {\bibinfo {volume} {102}},\ \bibinfo
  {pages} {217002} (\bibinfo {year} {2009})}\BibitemShut {NoStop}%
\bibitem [{\citenamefont {Kope{\'c}}\ and\ \citenamefont
  {Polak}(2000)}]{kopec2000}%
  \BibitemOpen
  \bibfield  {author} {\bibinfo {author} {\bibfnamefont {T.}~\bibnamefont
  {Kope{\'c}}}\ and\ \bibinfo {author} {\bibfnamefont {T.}~\bibnamefont
  {Polak}},\ }\href {\doibase 10.1103/PhysRevB.62.14419} {\bibfield  {journal}
  {\bibinfo  {journal} {Phys. Rev. B}\ }\textbf {\bibinfo {volume} {62}},\
  \bibinfo {pages} {14419} (\bibinfo {year} {2000})}\BibitemShut {NoStop}%
\bibitem [{\citenamefont {Mizokawa}\ \emph {et~al.}(2000)\citenamefont
  {Mizokawa}, \citenamefont {Khomskii},\ and\ \citenamefont
  {Sawatzky}}]{mizokawa2000}%
  \BibitemOpen
  \bibfield  {author} {\bibinfo {author} {\bibfnamefont {T.}~\bibnamefont
  {Mizokawa}}, \bibinfo {author} {\bibfnamefont {D.~I.}\ \bibnamefont
  {Khomskii}}, \ and\ \bibinfo {author} {\bibfnamefont {G.~A.}\ \bibnamefont
  {Sawatzky}},\ }\href {\doibase 10.1103/PhysRevB.61.11263} {\bibfield
  {journal} {\bibinfo  {journal} {Phys. Rev. B}\ }\textbf {\bibinfo {volume}
  {61}},\ \bibinfo {pages} {11263} (\bibinfo {year} {2000})}\BibitemShut
  {NoStop}%
\bibitem [{\citenamefont {Bisogni}\ \emph {et~al.}(2016)\citenamefont
  {Bisogni}, \citenamefont {Catalano}, \citenamefont {Green}, \citenamefont
  {Gibert}, \citenamefont {Scherwitzl}, \citenamefont {Huang}, \citenamefont
  {Strocov}, \citenamefont {Zubko}, \citenamefont {Balandeh}, \citenamefont
  {Triscone}, \citenamefont {Sawatzky},\ and\ \citenamefont
  {Schmitt}}]{bisogni2016ligandHole}%
  \BibitemOpen
  \bibfield  {author} {\bibinfo {author} {\bibfnamefont {V.}~\bibnamefont
  {Bisogni}}, \bibinfo {author} {\bibfnamefont {S.}~\bibnamefont {Catalano}},
  \bibinfo {author} {\bibfnamefont {R.~J.}\ \bibnamefont {Green}}, \bibinfo
  {author} {\bibfnamefont {M.}~\bibnamefont {Gibert}}, \bibinfo {author}
  {\bibfnamefont {R.}~\bibnamefont {Scherwitzl}}, \bibinfo {author}
  {\bibfnamefont {Y.}~\bibnamefont {Huang}}, \bibinfo {author} {\bibfnamefont
  {V.~N.}\ \bibnamefont {Strocov}}, \bibinfo {author} {\bibfnamefont
  {P.}~\bibnamefont {Zubko}}, \bibinfo {author} {\bibfnamefont
  {S.}~\bibnamefont {Balandeh}}, \bibinfo {author} {\bibfnamefont {J.-M.}\
  \bibnamefont {Triscone}}, \bibinfo {author} {\bibfnamefont {G.}~\bibnamefont
  {Sawatzky}}, \ and\ \bibinfo {author} {\bibfnamefont {T.}~\bibnamefont
  {Schmitt}},\ }\href {https://www.nature.com/articles/ncomms13017} {\bibfield
  {journal} {\bibinfo  {journal} {Nature Communications}\ }\textbf {\bibinfo
  {volume} {7}},\ \bibinfo {pages} {13017} (\bibinfo {year}
  {2016})}\BibitemShut {NoStop}%
\bibitem [{\citenamefont {Chakravarty}\ \emph {et~al.}(2004)\citenamefont
  {Chakravarty}, \citenamefont {Kee},\ and\ \citenamefont
  {V{\"o}lker}}]{chakravarty2004explanation}%
  \BibitemOpen
  \bibfield  {author} {\bibinfo {author} {\bibfnamefont {S.}~\bibnamefont
  {Chakravarty}}, \bibinfo {author} {\bibfnamefont {H.-Y.}\ \bibnamefont
  {Kee}}, \ and\ \bibinfo {author} {\bibfnamefont {K.}~\bibnamefont
  {V{\"o}lker}},\ }\href {https://www.nature.com/articles/nature02348}
  {\bibfield  {journal} {\bibinfo  {journal} {Nature}\ }\textbf {\bibinfo
  {volume} {428}},\ \bibinfo {pages} {53} (\bibinfo {year} {2004})}\BibitemShut
  {NoStop}%
\bibitem [{\citenamefont {Chu}\ \emph {et~al.}(2015)\citenamefont {Chu},
  \citenamefont {Deng},\ and\ \citenamefont {Lv}}]{chu2015hole}%
  \BibitemOpen
  \bibfield  {author} {\bibinfo {author} {\bibfnamefont {C.}~\bibnamefont
  {Chu}}, \bibinfo {author} {\bibfnamefont {L.}~\bibnamefont {Deng}}, \ and\
  \bibinfo {author} {\bibfnamefont {B.}~\bibnamefont {Lv}},\ }\href
  {https://www.sciencedirect.com/science/article/pii/S0921453415000878}
  {\bibfield  {journal} {\bibinfo  {journal} {Physica C: Superconductivity and
  its Applications}\ }\textbf {\bibinfo {volume} {514}},\ \bibinfo {pages}
  {290} (\bibinfo {year} {2015})}\BibitemShut {NoStop}%
\bibitem [{\citenamefont {Luo}\ \emph {et~al.}(2023{\natexlab{b}})\citenamefont
  {Luo}, \citenamefont {Chen}, \citenamefont {Li}, \citenamefont {Gao},
  \citenamefont {Yin}, \citenamefont {Yan}, \citenamefont {Miao}, \citenamefont
  {Luo}, \citenamefont {Shu}, \citenamefont {Chen}, \citenamefont {Lin},
  \citenamefont {Zhang}, \citenamefont {Wang}, \citenamefont {Zhang},
  \citenamefont {Yang}, \citenamefont {Peng}, \citenamefont {Liu},
  \citenamefont {Zhao}, \citenamefont {Xu}, \citenamefont {Xiang},\ and\
  \citenamefont {Zhou}}]{luo2023electronic}%
  \BibitemOpen
  \bibfield  {author} {\bibinfo {author} {\bibfnamefont {X.}~\bibnamefont
  {Luo}}, \bibinfo {author} {\bibfnamefont {H.}~\bibnamefont {Chen}}, \bibinfo
  {author} {\bibfnamefont {Y.}~\bibnamefont {Li}}, \bibinfo {author}
  {\bibfnamefont {Q.}~\bibnamefont {Gao}}, \bibinfo {author} {\bibfnamefont
  {C.}~\bibnamefont {Yin}}, \bibinfo {author} {\bibfnamefont {H.}~\bibnamefont
  {Yan}}, \bibinfo {author} {\bibfnamefont {T.}~\bibnamefont {Miao}}, \bibinfo
  {author} {\bibfnamefont {H.}~\bibnamefont {Luo}}, \bibinfo {author}
  {\bibfnamefont {Y.}~\bibnamefont {Shu}}, \bibinfo {author} {\bibfnamefont
  {Y.}~\bibnamefont {Chen}}, \bibinfo {author} {\bibfnamefont {C.}~\bibnamefont
  {Lin}}, \bibinfo {author} {\bibfnamefont {S.}~\bibnamefont {Zhang}}, \bibinfo
  {author} {\bibfnamefont {Z.}~\bibnamefont {Wang}}, \bibinfo {author}
  {\bibfnamefont {F.}~\bibnamefont {Zhang}}, \bibinfo {author} {\bibfnamefont
  {F.}~\bibnamefont {Yang}}, \bibinfo {author} {\bibfnamefont {Q.}~\bibnamefont
  {Peng}}, \bibinfo {author} {\bibfnamefont {G.}~\bibnamefont {Liu}}, \bibinfo
  {author} {\bibfnamefont {L.}~\bibnamefont {Zhao}}, \bibinfo {author}
  {\bibfnamefont {Z.}~\bibnamefont {Xu}}, \bibinfo {author} {\bibfnamefont
  {T.}~\bibnamefont {Xiang}}, \ and\ \bibinfo {author} {\bibfnamefont {X.~J.}\
  \bibnamefont {Zhou}},\ }\href
  {https://www.nature.com/articles/s41567-023-02206-0} {\bibfield  {journal}
  {\bibinfo  {journal} {Nature Physics}\ }\textbf {\bibinfo {volume} {19}},\
  \bibinfo {pages} {1841} (\bibinfo {year} {2023}{\natexlab{b}})}\BibitemShut
  {NoStop}%
\end{thebibliography}%

\appendix
\renewcommand{\theequation}{A\arabic{equation}}
\setcounter{equation}{0}
\renewcommand{\thefigure}{A\arabic{figure}}
\setcounter{figure}{0}
\renewcommand{\thetable}{A\arabic{table}}
\setcounter{table}{0}
\begin{widetext}
\section{Appendix A. Effective inter-layer $3d_{x^2-y^2}$-orbital spin-exchange interaction}
In the two orbital bilayer system,
the $3d_{z^2}$ orbitals in the two NiO$_2$ layers couple via the
hybridization with the O-2p orbitals in the intercalated LaO layer.
In the strong coupling limit, an effective nearest-neighbor antiferromagnetic (AFM) spin exchange $J_{\perp}$ between the two $3d_{z^2}$ electrons is generated {~\cite{ZhangGM2023DMRG,Yi_Feng2023}}.
For simplicity, we can focus on a pair of nearest-neighbor sites, $1$ and $2$, lying at the two layers.
The interacting spin Hamiltonian of the four spins in the two orbitals, $3d_{x^2-y^2}$ and $3d_{z^2}$, at the two sites $i=1,2$ is given by,
\begin{align}
&\hat{H}_{12}= -J_H \bm{S}_{z^2,1}\cdot\bm{S}_{x^2,1}
-J_H \bm{S}_{z^2,2}\cdot\bm{S}_{x^2,2}
+J_{\perp} \bm{S}_{z^2,1}\cdot\bm{S}_{z^2,2},
\label{eqA:spinH12}
\end{align}
with the on-site Hund's coupling $J_H$ and interlayer spin exchange $J_{\perp}$.
In the following, we will obtain an effective interlayer spin-exchange interaction between the two nearest-neighbor $3d_{z^2-y^2}$ orbitals at the two layers from $\hat{H}_{12}$.

In the spin-coherent state path integral\cite{auerbach1998},
we treat $\bm{S}_{z^2,1}$ and $\bm{S}_{z^2,2}$ as the basic field variables and integrate out them to obtain an effective spin interaction between the two $3d_{x^2-y^2}$ orbitals.
The partition function of the spin system Eq.~(\ref{eqA:spinH12}) is given by \cite{auerbach1998},
\begin{align}
Z=& \mathrm{Tr}T_{\tau} \Big( \exp\big[ -\int_0^{\beta}d\tau \hat{H}(\tau) \big] \Big)
={\int \mathcal{D} \hat{\Omega}_{z^2} \mathcal{D} \hat{\Omega}_{x^2}
\exp\big( -\tilde{S}[\hat{\Omega}_{z^2},\hat{\Omega}_{x^2}] \big)},
\label{eqA:partitionH12}
\end{align}
where {$\bm{S}_{z^2,i}\rightarrow S\hat{\Omega}_{z^2,i}$($i=1,2$, $S=1/2$) and similar for the $3d_{x^2}$ orbital.}
The spin action in the imaginary time formulation is given by,
{
\begin{align*}
\tilde{S}[\hat{\Omega}_{z^2},\hat{\Omega}_{x^2}]
=&\tilde{S}_{\omega}[\hat{\Omega}_{z^2}]
+\tilde{S}_{\omega}[\hat{\Omega}_{x^2}]
+\tilde{S}_{\perp}[\hat{\Omega}_{z^2}]
+\tilde{S}_{H}[\hat{\Omega}_{z^2},\hat{\Omega}_{x^2}],  \\
\tilde{S}_{\perp}[\hat{\Omega}_{z^2}]
=&S^2 J_{\perp} \int_0^{\beta} d\tau \hat{\Omega}_{z^2,1} \cdot\hat{\Omega}_{z^2,2},	\\
\tilde{S}_{\omega}[\hat{\Omega}_{z^2}]=& iS \sum_{i=1,2} \int_0^{\beta} d\tau (\partial_{\tau}\phi_{z^2,i}) \cos\theta_{z^2,i},\qquad
\tilde{S}_{\omega}[\hat{\Omega}_{x^2}]= iS \sum_{i=1,2} \int_0^{\beta} d\tau (\partial_{\tau}\phi_{x^2,i}) \cos\theta_{x^2,i},	\\
\tilde{S}_H[\hat{\Omega}_{z^2},\hat{\Omega}_{x^2}]=& -S^2 J_H \int_0^{\beta} d\tau \Big(
\hat{\Omega}_{x^2,1} \cdot \hat{\Omega}_{z^2,1}
+ \hat{\Omega}_{x^2,2} \cdot \hat{\Omega}_{z^2,2} \Big).
\end{align*} }
Here, $\tilde{S}_{\omega}$ is the Berry phase contribution, $\tilde{S}_{\perp}$ is the interlayer AFM spin exchange of the $3d_{z^2}$ orbitals and $\tilde{S}_H$ is the on-site Hund's coupling.
{The unit vector for the $3d_{z^2}$ spin is parameterized as $\hat{\Omega}_{z^2,i} =\big( \sin\theta_{z^2,i}\cos\phi_{z^2,i}, \sin\theta_{z^2,i} \sin \phi_{z^2,i}, \cos\theta_{z^2,i}\big)$,
with a boundary condition of $\hat{\Omega}_{z^2,i}(\beta)=\hat{\Omega}_{z^2,i}(0)$.
A similar paramaterization is also applied to the $3d_{x^2-y^2}$ orbital spins.}

{
In the strong Hund's coupling limit ($J_H\gg J_{\perp}$), the on-site two orbitals exhibit a tendency to form a spin triplet configuration.
Consequently, the spin orientations $(\hat{\Omega}_{z^2,1},\hat{\Omega}_{z^2,2})$ exhibit a preference to align with the directions of $(\hat{\Omega}_{x^2,1},\hat{\Omega}_{x^2,2})$,
\begin{align}
\hat{\Omega}_{z^2,1}(\tau)\rightarrow \hat{\Omega}_{x^2,1}(\tau),\qquad
\hat{\Omega}_{z^2,2}(\tau)\rightarrow \hat{\Omega}_{x^2,2}(\tau).
\label{eqA:averageSpinOmega}
\end{align}
By integrating out the spin degrees of freedom associated with the $3d_{z^2}$-orbitals,
we can effectively substitute the unit spin vectors $\hat{\Omega}_{z^2,i}$ by $\hat{\Omega}_{x^2,i}$ in the partition function Eq.~(\ref{eqA:partitionH12}),
\begin{align}
Z=&\int \mathcal{D} \hat{\Omega}_{x^2} \exp\big( -\tilde{S}_{\omega}[\hat{\Omega}_{x^2}] \big)
\int \mathcal{D} \hat{\Omega}_{z^2} \exp\big( -\tilde{S}_{\perp}[\hat{\Omega}_{z^2}] -\tilde{S}_{\omega}[\hat{\Omega}_{z^2}] -\tilde{S}_{H}[\hat{\Omega}_{z^2},\hat{\Omega}_{x^2}] \big)    \notag\\
\rightarrow&\int \mathcal{D} \hat{\Omega}_{x^2} \exp\big( -2\tilde{S}_{\omega}[\hat{\Omega}_{x^2}] -\tilde{S}_{\perp}[\hat{\Omega}_{x^2}] \big).
\label{eqA:PerturZ}
\end{align}
In the semi-classical approximation, an effective AFM spin-exchange interaction between $3d_{x^2-y^2}$ orbitals is generated,
\begin{align*}
\tilde{S}_{\perp}[\hat{\Omega}_{x^2}]
=J_{\perp} S^2 \int_0^{\beta} d\tau \hat{\Omega}_{x^2,1}\cdot \hat{\Omega}_{x^2,2}
=J_{\perp} \int_0^{\beta} d\tau {\bm{S}}_{x^2,1}\cdot{\bm{S}}_{x^2,2}.
\end{align*}
Therefore, the two-orbital model reduces to a single $3d_{x^2-y^2}$ orbital model, featuring an effective interlayer AFM spin exchange between the two $3d_{x^2-y^2}$ orbitals situated in the two layers.
}

\section{Appendix B. Hamiltonian operator method under Hund's rule}
In this section, we consider {a} direct Hamiltonian operator formulation to {establish} the equivalence between the exchange operators $\bm{S}_{x^2,1}\cdot\bm{S}_{x^2,2}$ and $\bm{S}_{z^2,1}\cdot\bm{S}_{z^2,2}$.
{Similar to} before, {we} consider a pair of nearest-neighbor sites, $1$ and $2$, lying at the two layers.
There are totally four spins and the Hilbert space { encompasses} $2^4=16$ states.
Impose the Hund's rule for $3d_{z^2}$ and $3d_{x^2-y^2}$ orbitals described by the {strong} Hund's couplings,
\begin{align}
\hat{H}_{\text{Hund}} =
-J_H \bm{S}_{z^2,1}\cdot \bm{S}_{x^2,1}
-J_H \bm{S}_{z^2,2}\cdot \bm{S}_{x^2,2},
\label{eqB:HundsCouplnig}
\end{align}
the two spins at a given site should form a spin-triplet (spin-$1$) state.
The Hilbert space reduces to the following $9=3\times 3$ physical states under the Hund's rule,
\begin{align*}
\begin{cases}
|+1\rangle_1=\displaystyle\big| \underset{d_{z^2}}{\uparrow},\ \underset{d_{x^2}}{\uparrow}\big\rangle_{1}	\\
\\
|0\rangle_1=\displaystyle\frac{1}{\sqrt{2}}
\Big(\big| \underset{d_{z^2}}{\uparrow},\ \underset{d_{x^2}}{\downarrow}\big\rangle_{1}
+\big| \underset{d_{z^2}}{\downarrow},\ \underset{d_{x^2}}{\uparrow}\big\rangle_{1}
\Big)	\\
\\
|-1\rangle_1=\displaystyle\big| \underset{d_{z^2}}{\downarrow},\ \underset{d_{x^2}}{\downarrow}\big\rangle_{1}
\end{cases}
\otimes
\begin{cases}
|+1\rangle_2=\displaystyle\big| \underset{d_{z^2}}{\uparrow},\ \underset{d_{x^2}}{\uparrow}\big\rangle_{2}	\\
\\
|0\rangle_2=\displaystyle\frac{1}{\sqrt{2}}
\Big(\big| \underset{d_{z^2}}{\uparrow},\ \underset{d_{x^2}}{\downarrow}\big\rangle_{2}
+\big| \underset{d_{z^2}}{\downarrow},\ \underset{d_{x^2}}{\uparrow}\big\rangle_{2}
\Big)	\\
\\
|-1\rangle_2=\displaystyle\big| \underset{d_{z^2}}{\downarrow},\ \underset{d_{x^2}}{\downarrow}\big\rangle_{2}
\end{cases}
\end{align*}
where $|\sigma_{z^2},\sigma_{x^2}\rangle_i$ represents the spin configuration of the two $E_g$-orbitals in the site $i$.
These $9$ states are eigenstates of spin-$1$ operators,
$\bm{S}_1=\bm{S}_{x^2,1}+\bm{S}_{z^2,1},\bm{S}_2=\bm{S}_{x^2,2}+\bm{S}_{z^2,2}$
with $S_1=S_2=1$.
Their energies under Hund's coupling Eq.~(\ref{eqB:HundsCouplnig}) are degenerate, $E_{\text{Hund}}=-\frac{J_H}{2}$.
The $9$ combinations of two spin-triplet states can form total spin $J=0,1,2$ states, with total spin operator
$\bm{J}=\bm{S}_1+\bm{S}_2=\bm{S}_{x^2,1}+\bm{S}_{z^2,1}+\bm{S}_{x^2,2}+\bm{S}_{z^2,2}$.
For total spin-$2$, there are five states, corresponding to $M=\pm 2,\pm 1,0$,
\begin{align*}
|\underset{J}{2},\underset{M}{+2}\rangle =&|+1\rangle_1 |+1\rangle_2
=|\uparrow\uparrow\rangle_1 |\uparrow\uparrow\rangle_2	\\
|\underset{J}{2},\underset{M}{+1}\rangle
=&\frac{1}{\sqrt{2}}\Big(|+1\rangle_1 |0\rangle_2 +|0\rangle_1 |+1\rangle_2\Big)
=\frac{1}{\sqrt{2}}\Big(|\uparrow\uparrow\rangle_1 |0\rangle_2 +|0\rangle_1 |\uparrow\uparrow\rangle_2\Big)	\\
|\underset{J}{2},\underset{M}{0}\rangle
=&\frac{1}{\sqrt{6}}\Big(|+1\rangle_1 |-1\rangle_2
+2|0\rangle_1 |0\rangle_2 +|-1\rangle_1 |+1\rangle_2\Big)	\\
=&\frac{1}{\sqrt{6}}\Big(|\uparrow\uparrow\rangle_1 |\downarrow\downarrow\rangle_2
+2|0\rangle_1 |0\rangle_2 +|\downarrow\downarrow\rangle_1 |\uparrow\uparrow\rangle_2\Big)	\\
|\underset{J}{2},\underset{M}{-1}\rangle
=&\frac{1}{\sqrt{2}}\Big(|-1\rangle_1 |0\rangle_2 +|0\rangle_1 |-1\rangle_2\Big)
=\frac{1}{\sqrt{2}}\Big(|\downarrow\downarrow\rangle_1 |0\rangle_2 +|0\rangle_1 |\downarrow\downarrow\rangle_2\Big)		\\
|\underset{J}{2},\underset{M}{-2}\rangle =&|-1\rangle_1 |-1\rangle_2
=|\downarrow\downarrow\rangle_1 |\downarrow\downarrow\rangle_2.
\end{align*}
For total spin-$1$, there are three states,
corresponding to $M=\pm 1,0$,
\begin{align*}
|\underset{J}{1},\underset{M}{+1}\rangle
=&\frac{1}{\sqrt{2}}\Big(|+1\rangle_1 |0\rangle_2 -|0\rangle_1 |+1\rangle_2\Big)
=\frac{1}{\sqrt{2}}\Big(|\uparrow\uparrow\rangle_1 |0\rangle_2 -|0\rangle_1 |\uparrow\uparrow\rangle_2\Big),\\
|\underset{J}{1},\underset{M}{0}\rangle
=&\frac{1}{\sqrt{2}}\Big(|+1\rangle_1 |-1\rangle_2
-|-1\rangle_1 |+1\rangle_2\Big)
=\frac{1}{\sqrt{2}}\Big(|\uparrow\uparrow\rangle_1 |\downarrow\downarrow\rangle_2 -|\downarrow\downarrow\rangle_1 |\uparrow\uparrow\rangle_2\Big),	\\
|\underset{J}{1},\underset{M}{-1}\rangle
=&\frac{1}{\sqrt{2}}\Big(|-1\rangle_1 |0\rangle_2 -|0\rangle_1 |-1\rangle_2\Big)
=\frac{1}{\sqrt{2}}\Big(|\downarrow\downarrow\rangle_1 |0\rangle_2 -|0\rangle_1 |\downarrow\downarrow\rangle_2\Big).
\end{align*}
For total spin-$0$, there is only one state,
\begin{align*}
|\underset{J}{0},\underset{M}{0}\rangle =&\frac{1}{\sqrt{3}}\Big(
|+1\rangle_1 |-1\rangle_2
-|0\rangle_1 |0\rangle_2
+|-1\rangle_1 |+1\rangle_2\Big)		\\
=&\frac{1}{\sqrt{3}} \big(|\uparrow\uparrow\rangle_1|\downarrow\downarrow\rangle_2
-|0\rangle_1 |0\rangle_2
+|\downarrow\downarrow\rangle_1|\uparrow\uparrow\rangle_2\big).
\end{align*}
These $9$ states are symmetric under exchanging the spins of two orbitals at a given site.

Next, we consider the action of the following spin-exchange operators on these $9$ states,
\begin{align*}
\hat{H}_{z^2}
=& \bm{S}_{z^2,1}\cdot\bm{S}_{z^2,2}
= {S}_{z^2,1}^{z}{S}_{z^2,2}^{z}
+ \frac{1}{2} \Big({S}_{z^2,1}^+{S}_{z^2,2}^-
+{S}_{z^2,1}^-{S}_{z^2,2}^+\Big),	\\
\hat{H}_{x^2}
=&\bm{S}_{x^2,1}\cdot\bm{S}_{x^2,2},\qquad
\hat{H}_{x^2,z^2}
=\bm{S}_{x^2,1}\cdot\bm{S}_{z^2,2},\qquad
\hat{H}_{z^2,x^2}
=\bm{S}_{z^2,1}\cdot\bm{S}_{x^2,2}.
\end{align*}
We will focus on $\hat{H}_{z^2}$ and the others are similar.
The action of $\hat{H}_{z^2}$ on the five total spin-$2$ states is,
\begin{align*}
\hat{H}_{z^2}|J={2},M\rangle
=\frac{1}{4} |J={2},M\rangle.
\end{align*}
These five states ($M=\pm 2,\pm 1,0$) are eigenstates of $\hat{H}_{z^2}$.
Next, consider the three total spin-$1$ states.
We can obtain the following equations,
\begin{align*}
\Big( \hat{H}_{z^2}
+\frac{1}{4} \Big)|\underset{J}{1},\underset{M}{+1}\rangle
=&\frac{1}{2\sqrt{2}} \Big(|\uparrow\uparrow\rangle_1|s\rangle_2
-|s\rangle_1 |\uparrow\uparrow\rangle_2 \Big), \\
\Big( \hat{H}_{z^2}
+\frac{1}{4} \Big)|\underset{J}{1},\underset{M}{0}\rangle
=&\frac{1}{2\sqrt{2}}
\Big(|0\rangle_1|s\rangle_2 -|s\rangle_1|0\rangle_2 \Big),
\end{align*}
where $|s\rangle_i$ ($i=1,2$) is spin-singlet configuration at the site $i$.
The total spin-$1$ states are not the exact eigenstates of $\hat{H}_{z^2}$.
However, when restricted to the physical Hilbert space under the Hund's rule,
the total spin-$1$ states are approximately eigenstates,
\begin{align*}
\hat{H}_{z^2}|J=1,{M}\rangle
+\frac{1}{4}|J=1,{M}\rangle  \sim 0,\qquad
\hat{H}_{z^2}|J=1,{M}\rangle
\simeq -\frac{1}{4}|J=1,{M}\rangle.
\end{align*}
Finally, consider the total spin-$0$ case,
\begin{align*}
&\Big(\hat{H}_{z^2}+\frac{1}{2}\Big)|0,0\rangle
=\frac{\sqrt{3}}{4} |s\rangle_1 |s\rangle_2,
\end{align*}
which does not belong to the physical Hilbert spcae under the Hund's rule.
For the total spin-$0$, we can have
\begin{align*}
\hat{H}_{z^2}|0,0\rangle
\simeq -\frac{1}{2}|0,0\rangle.
\end{align*}

In summary, the $9$ physical states {represent} eigenstates of $\hat{H}_{z^2}$ {within} the restricted physical Hilbert space under the Hund's rule.
Since these $9$ states are symmetric in the $3d_{z^2}$ and $3d_{x^2}$ orbital, this argument holds for $\hat{H}_{x^2}$.
For the four spin-exchange combinations,
\begin{align*}
\hat{H}_{z^2}
=& \bm{S}_{z^2,1}\cdot\bm{S}_{z^2,2},\qquad
\hat{H}_{x^2}
=\bm{S}_{x^2,1}\cdot\bm{S}_{x^2,2},\qquad
\hat{H}_{x^2,z^2}
=\bm{S}_{x^2,1}\cdot\bm{S}_{z^2,2},\qquad
\hat{H}_{z^2,x^2}
=\bm{S}_{z^2,1}\cdot\bm{S}_{x^2,2},
\end{align*}
they are equivalent {within} the physical Hilbert space {according to} the Hund's rule, as summarised in Tab.~(\ref{tab:SpinExchangeHunds}).
Here, AFM spin-$1$ exchange ($\bm{S}_{1}\cdot\bm{S}_{2}$) between the two spin-triplet states at the two sites is a summation of these four equivalent spin-$\frac{1}{2}$ exchanges.

\begin{table}
\centering
\begin{tabular}{|c|c|c|c|}
\hline
 & \makecell{$\qquad$total $\qquad$\\ spin-$0$} & \makecell{$\qquad$total $\qquad$\\ spin-$1$} & \makecell{$\qquad$total $\qquad$\\ spin-$2$} \\
\hline
\makecell{$\phantom{a}$ \\ $\bm{S}_{z^2,1}\cdot\bm{S}_{z^2,2}$ \\ $\phantom{a}$} & $-\displaystyle \frac{1}{2}$ & $-\displaystyle\frac{1}{4}$ & $\displaystyle\frac{1}{4}$ \\
\hline
\makecell{$\phantom{a}$ \\ $\bm{S}_{x^2,1}\cdot\bm{S}_{x^2,2}$ \\ $\phantom{a}$} & $-\displaystyle \frac{1}{2}$ & $-\displaystyle\frac{1}{4}$ & $\displaystyle\frac{1}{4}$ \\
\hline
\makecell{$\phantom{a}$ \\ $\bm{S}_{x^2,1}\cdot\bm{S}_{z^2,2}$ \\ $\phantom{a}$} & $-\displaystyle \frac{1}{2}$ & $-\displaystyle\frac{1}{4}$ & $\displaystyle\frac{1}{4}$ \\
\hline
\makecell{$\phantom{a}$ \\ $\bm{S}_{z^2,1}\cdot\bm{S}_{x^2,2}$ \\ $\phantom{a}$} & $-\displaystyle \frac{1}{2}$ & $-\displaystyle\frac{1}{4}$ & $\displaystyle\frac{1}{4}$ \\
\hline
\makecell{$\phantom{a}$ \\$\bm{S}_{1}\cdot\bm{S}_{2}$ \\ $\phantom{a}$} & $-\displaystyle 2$ & $-\displaystyle1$ & $\displaystyle 1$ \\
\hline
\end{tabular}
\caption{Eigenvalues of several spin-exchange interactions for the $9$ states under Hund's rule.
The spin-$1$ operators are $\bm{S}_1=\bm{S}_{x^2,1}+\bm{S}_{z^2,1}$ and
$\bm{S}_2=\bm{S}_{x^2,2}+\bm{S}_{z^2,2}$}
\label{tab:SpinExchangeHunds}
\end{table}

\section{Appendix C. Slave-boson mean-field approach}
\label{SLPapproach}
In this section, we provide the explicit details of the slave-particle approach \cite{kotliar1988,lee2006htsc} {employed in the main text}.
The Hamiltonian of the single-orbital bilayer $t$-$J$ model is given by,
\begin{equation}
\begin{aligned}
&H
=-t \sum_{\langle \bm{i},\bm{j} \rangle,\alpha,\sigma}
\big( {c}^{\dagger}_{\bm{i}\alpha\sigma}{c}_{\bm{j}\alpha\sigma}+\text{h.c.} \big)
+J_{\parallel} \sum_{\langle \bm{i},\bm{j} \rangle,\alpha} \bm{S}_{\bm{i}\alpha}\cdot\bm{S}_{\bm{j}\alpha}
-t_{\perp} \sum_{\bm{i}\sigma}
\big( {c}^{\dagger}_{\bm{i}1\sigma}{c}_{\bm{i}2\sigma}+\text{h.c.} \big)
+J_{\perp} \sum_{\bm{i}} \bm{S}_{\bm{i}1}\cdot \bm{S}_{\bm{i}2}.
\end{aligned}
\end{equation}
In the slave-boson mean field theory, the electron operator is expressed as $c_{\bm{i}\alpha\sigma}^{\dagger}=f_{\bm{i}\alpha\sigma}^{\dagger}b_{\bm{i}\alpha}$, where $f_{\bm{i}\alpha\sigma}^{\dagger}/b_{\bm{i}\alpha}$ is the spinon/holon creation/annihilation operator,
with the local constraint $\sum_{\sigma}f_{\bm{i}\alpha\sigma}^{\dagger} f_{\bm{i}\alpha\sigma} +b_{\bm{i}\alpha}^{\dagger} b_{\bm{i}\alpha} =1$.
Introduce the intra-layer hoppings and pairings and the interlayer ones,
\begin{equation}
\begin{aligned}
\chi_{\bm{i}\bm{j}}^{(\alpha)}
=&f_{\bm{j}\alpha\uparrow}^{\dagger} f_{\bm{i}\alpha\uparrow}
+f_{\bm{j}\alpha\downarrow}^{\dagger} f_{\bm{i}\alpha\downarrow},\qquad
\Delta_{\bm{i}\bm{j}}^{(\alpha)}
=f_{\bm{j}\alpha\downarrow} f_{\bm{i}\alpha\uparrow }
-f_{\bm{j}\alpha\uparrow} f_{\bm{i}\alpha\downarrow }, \\
\chi_{\bm{i},\perp}
=&f_{\bm{i}2\uparrow}^{\dagger} f_{\bm{i}1\uparrow}
+f_{\bm{i}2\downarrow}^{\dagger} f_{\bm{i}1\downarrow},\qquad
\Delta_{\bm{i},\perp}
=f_{\bm{i}2\downarrow } f_{\bm{i}1\uparrow}
-f_{\bm{i}2\uparrow} f_{\bm{i}1\downarrow},
\end{aligned}
\end{equation}
the Hamiltonian can be decoupled in the form,
\begin{align*}
H=&-t \sum_{\langle \bm{i},\bm{j} \rangle,\alpha}
\Big( {b}_{\bm{i}\alpha} b^{\dagger}_{\bm{j}\alpha}
\sum_{\sigma} {f}^{\dagger}_{\bm{i}\alpha\sigma} {f}_{\bm{j}\alpha\sigma} +\text{h.c.} \Big)
-t_{\perp} \sum_{\bm{i}\sigma}
\Big( {b}_{\bm{i}1} b^{\dagger}_{\bm{i}2}
\sum_{\sigma} {f}^{\dagger}_{\bm{i}1\sigma} {f}_{\bm{i}2\sigma} +\text{h.c.} \Big)  \\
&+\sum_{\bm{i}\alpha} \lambda_{\bm{i}\alpha}\big(\sum_{\sigma}f_{\bm{i}\alpha\sigma}^{\dagger} f_{\bm{i}\alpha\sigma} +b_{\bm{i}\alpha}^{\dagger} b_{\bm{i}\alpha}-1\big)  \\
&-\frac{3J_{\parallel}}{8} \sum_{\langle \bm{i},\bm{j} \rangle,\alpha} \Big[ \chi_{\bm{i}\bm{j}}^{(\alpha)}
\big(f_{\bm{i}\alpha\uparrow}^{\dagger} f_{\bm{j}\alpha\uparrow}
+f_{\bm{i}\alpha\downarrow}^{\dagger} f_{\bm{j}\alpha\downarrow}\big) +\text{h.c.} -|\chi_{\bm{i}\bm{j}}^{(\alpha)}|^2  \Big] \\
&-\frac{3J_{\parallel}}{8} \sum_{\langle \bm{i},\bm{j} \rangle,\alpha} \Big[ \Delta_{\bm{i}\bm{j}}^{(\alpha)}
\big(f_{\bm{i}\alpha\uparrow}^{\dagger}f_{\bm{j}\alpha\downarrow}^{\dagger}
-f_{\bm{i}\alpha\downarrow}^{\dagger}f_{\bm{j}\alpha\uparrow}^{\dagger} \big) +\text{h.c.} -|\Delta_{\bm{i}\bm{j}}^{(\alpha)}|^2  \Big]   \\
&-\frac{3J_{\perp}}{8} \sum_{\bm{i}} \Big[ \chi_{\bm{i},\perp}
\big(f_{\bm{i}1\uparrow}^{\dagger} f_{\bm{i}2\uparrow}
+f_{\bm{i}1\downarrow}^{\dagger} f_{\bm{i}2\downarrow}\big) +\text{h.c.} -|\chi_{\bm{i},\perp}|^2  \Big] \\
&-\frac{3J_{\perp}}{8} \sum_{\bm{i}} \Big[ \Delta_{\bm{i},\perp}
\big(f_{\bm{i}1\uparrow}^{\dagger}f_{\bm{i}2\downarrow}^{\dagger}
-f_{\bm{i}1\downarrow}^{\dagger}f_{\bm{i}2\uparrow}^{\dagger} \big) +\text{h.c.} -|\Delta_{\bm{i},\perp}|^2  \Big].
\end{align*}
In the mean-field analysis, hoppings $\chi$, pairings $\Delta$ and Lagrange multipliers $\lambda_{\bm{i}\alpha}$ are replaced by their site-independent mean-field values,
\begin{align*}
\chi_{\bm{i}\bm{j}}^{(\alpha)} = \chi_{\bm{j}-\bm{i}}^{(\alpha)},\qquad
\chi_{\bm{i},\perp} = \chi_{z},\qquad
\Delta_{\bm{i}\bm{j}}^{(\alpha)} = \Delta_{\bm{j}-\bm{i}}^{(\alpha)},\qquad
\Delta_{\bm{i},\perp} = \Delta_{z},\qquad
\lambda_{\bm{i}\alpha}=\lambda.
\end{align*}
The holon is condensed ${b}_{\bm{i}\alpha}=\sqrt{\delta}$.
The mean-field Hamiltonian for the spinon part is given by,
\begin{align*}
&H_{f,\text{MF}}=\sum_{\bm{k}\alpha\sigma} \varepsilon_{\bm{k},\alpha} f_{\bm{k}\alpha\sigma}^{\dagger} f_{\bm{k}\alpha\sigma}
-\sum_{\bm{k}}
\Big[\Big(\frac{3}{8} J_{\perp} \chi_{z}+t_{\perp} \delta \Big)  \big(f_{\bm{k}1\uparrow}^{\dagger} f_{\bm{k}2\uparrow}
+f_{-\bm{k}1\downarrow}^{\dagger} f_{-\bm{k}2\downarrow} \big)+\text{h.c.} \Big]
\\
&+\sum_{\bm{k}\alpha} \big(F_{\bm{k},\alpha} f_{\bm{k}\alpha\uparrow}^{\dagger} f_{-\bm{k}\alpha\downarrow}^{\dagger} +\text{h.c.} \big)
-\frac{3}{8} J_{\parallel} \sum_{\bm{k}}
\Delta_{z} \Big[ \big(f_{\bm{k}1\uparrow}^{\dagger} f_{-\bm{k}2\downarrow}^{\dagger}
-f_{-\bm{k}1\downarrow}^{\dagger} f_{\bm{k}2\uparrow}^{\dagger} \big) +\text{h.c.} \Big] \\
&- \frac{3}{8} J_{\parallel} N \sum_{\alpha} \big(|\chi_{x}^{(\alpha)}|^2 +|\chi_{y}^{(\alpha)}|^2
+|\Delta_{x}^{(\alpha)}|^2 +|\Delta_{y}^{(\alpha)}|^2 \big)
+\frac{3}{8} J_{\perp} N \big(|\chi_{\perp}|^2 +|\Delta_{\perp}|^2 \big)
\end{align*}
where the intralayer kinectic energy and pairing are,
\begin{align*}
\varepsilon_{\bm{k},\alpha} =& -\frac{3}{8} J_{\parallel} \big[\chi_{x}^{(\alpha)} e^{-ik_x}
+\chi_{y}^{(\alpha)} e^{-ik_y} +\text{h.c.}\big]
-2t\delta \big[ \cos k_x +\cos k_y \big]  -\mu_f,    \\
F_{\bm{k},\alpha} =& -\frac{3}{4} J_{\parallel} \big[\Delta_{x}^{(\alpha)} \cos k_x +\Delta_{y}^{(\alpha)} \cos k_y \big],
\end{align*}
and $\mu_f$ is the chemical potential of the spinon field.
Introduce the Nambu spinor,
\begin{align*}
\psi_{\bm{k}}^{\dagger}
=\begin{pmatrix}
f_{\bm{k}1\uparrow}^{\dagger} &
f_{\bm{k}2\uparrow}^{\dagger} &
f_{-\bm{k}1\downarrow} &
f_{-\bm{k}2\downarrow}
\end{pmatrix},
\end{align*}
the mean-field Hamiltonian can be written as,
\begin{align*}
&H_{f,\text{MF}}
=\sum_{\bm{k}}
\psi_{\bm{k}}^{\dagger}
\begin{pmatrix}
\varepsilon_{\bm{k},1} & -\displaystyle \frac{3}{8} J_{\perp} \chi_{z} -t_{\perp} \delta & F_{\bm{k},1} & \displaystyle -\frac{3}{8} J_{\perp} \Delta_{z} \\
-\displaystyle \frac{3}{8} J_{\perp} \chi_{z}^* -t_{\perp} \delta & \varepsilon_{\bm{k},2} & \displaystyle -\frac{3}{8} J_{\perp} \Delta_{z} & F_{\bm{k},2} \\
F_{\bm{k},1}^* & \displaystyle -\frac{3}{8} J_{\perp} \Delta_{z}^* & -\varepsilon_{-\bm{k},1} & \displaystyle \frac{3}{8} J_{\perp} \chi_{z}^* +t_{\perp} \delta \\
\displaystyle -\frac{3}{8} J_{\perp}\Delta_{z}^* & F_{\bm{k},2}^* & \displaystyle \frac{3}{8} J_{\perp} \chi_{z} +t_{\perp} \delta & -\varepsilon_{-\bm{k},2}
\end{pmatrix}
\psi_{\bm{k}}   \\
&+\sum_{\bm{k}\alpha} \varepsilon_{-\bm{k},\alpha}
+\frac{3}{8} J_{\parallel} N \sum_{\alpha} \big(|\chi_{x}^{(\alpha)}|^2 +|\chi_{y}^{(\alpha)}|^2
+|\Delta_{x}^{(\alpha)}|^2 +|\Delta_{y}^{(\alpha)}|^2 \big)
+\frac{3}{8} J_{\perp} N \big(|\chi_{z}|^2 +|\Delta_{z}|^2 \big).
\end{align*}
The quadratic term in $H_{f,\text{MF}}$ can be diagonalized through a unitary transformation,
\begin{align*}
H_{f,\text{MF}}
=\sum_{\bm{k}} \psi_{\bm{k}}^{\dagger} H_{\bm{k}} \psi_{\bm{k}} +\cdots
=\sum_{\bm{k}} \gamma_{\bm{k}}^{\dagger} E_{\bm{k}} \gamma_{\bm{k}} +\cdots
\end{align*}
where $E_{\bm{k}}=\mathrm{diag}(E_{1\bm{k}},E_{2\bm{k}},E_{3\bm{k}},E_{4\bm{k}})$ is the eigenvalue matrix for $H_{\bm{k}}$ and $\gamma_{\bm{k}}^{\dagger}=(\gamma_{1\bm{k}}^{\dagger},\gamma_{2\bm{k}}^{\dagger},\gamma_{3\bm{k}}^{\dagger},\gamma_{4\bm{k}}^{\dagger})$ is the quasi-particle creation operator,
with a unitary transformation $U_{\bm{k}}$ diagonalizing $H_{f,\text{MF}}$: $H_{\bm{k}} =U_{\bm{k}} E_{\bm{k}} U_{\bm{k}}^{\dagger}$,
$\gamma_{\bm{k}}=U_{\bm{k}}^{\dagger} \psi_{\bm{k}}$.
Due to the particle-hole symmetry,
the quasi-particle spectrum can be arranged in the way
\begin{align*}
E_{1\bm{k}}=E_{1,\bm{k}}^{(+)}>0,\qquad
E_{2\bm{k}}=E_{2,\bm{k}}^{(+)}>0,\qquad
E_{3\bm{k}}=-E_{1,-\bm{k}}^{(+)}<0,\qquad
E_{4\bm{k}}=-E_{2,-\bm{k}}^{(+)}<0.
\end{align*}
and the free energy is
\begin{align*}F=&
-\frac{2}{\beta} \sum_{\bm{k},a=1,2} \ln\Big( 1+e^{-\beta E_{a\bm{k}}^{(+)}}\Big)
-\sum_{\bm{k},a=1,2} E_{a\bm{k}}^{(+)} -2\mu_f N	\\
& + \frac{3}{8} J_{\parallel} N \sum_{\alpha} \big(|\chi_{x}^{(\alpha)}|^2 +|\chi_{y}^{(\alpha)}|^2
+|\Delta_{x}^{(\alpha)}|^2 +|\Delta_{y}^{(\alpha)}|^2 \big)
+\frac{3}{8} J_{\perp} N \big(|\chi_{z}|^2 +|\Delta_{z}|^2 \big).
\end{align*}

\section*{Appendix D. High-$T_c$ SC driven by interlayer coupling}
In the slave-boson mean-field theory (SBMFT), the critical temperature $T_c$ for superconductivity is governed by the lower of two  temperatures: $T_{\text{BEC}}$ and $T_{\text{pair}}$ \cite{kotliar1988,lee2006htsc}.
Here, $T_{\text{BEC}}$ signifies the Bose-Einstein condensate temperature of holons, while $T_{\text{pair}}$ represents the pairing temperature of the spinons.
As the hole-doping level increases, $T_{\text{BEC}}$ becomes considerably larger than $T_{\text{pair}}$ \cite{kotliar1988}.
Consequently, for $x$ close to $0.25$, $T_c$ is predominantly dictated by $T_{\text{pair}}$.
The spinon pairing temperature $T_{\text{pair}}$ is derived by solving the finite-temperature mean-field gap equation self-consistently.
Fig.~\ref{fig:JperpTcFilling}(a) illustrates the filling level $x$ dependence of the deduced superconducting $T_c$ for various ratios of super-exchange strengths $J_{\perp}/J_{\parallel}$, with fixed $J_{\parallel}=0.4t$,
exhibiting a similar trend to the ground state superconducting order parameter depicted in Fig.3(a) in the main text.
Evidently, $T_c$ increases markedly with the enhancement of $x$ for all $J_{\perp}/J_{\parallel}$ ratios.

To delve deeper into the impact of interlayer coupling, Fig.\ref{fig:JperpTcFilling}(b) presents $T_c$ as a function of $J_{\perp}/J_{\parallel}$ for different fillings ($x=0.25,0.28,0.3,0.32$).
Remarkably, across all the experimentally relevant fillings, $T_c$ consistently and significantly increases with the rise of $J_{\perp}/J_{\parallel}$, particularly when $J_{\perp}>J_{\parallel}$.
The realistic value of $J_{\perp}/J_{\parallel}$ in La$_3$Ni$_2$O$_7$ (LNO) under pressure can be estimated from density functional theory (DFT) calculations \cite{YaoDX2023}, approximating $J_{\perp}/J_{\parallel}$ as $(0.635/0.48)^2 \approx 1.75$.
In Fig.\ref{fig:JperpTcFilling}(c), a comparative analysis of the filling dependence of $T_c$ between $J_{\perp}=0$ and the realistic $J_{\perp}$ is depicted.
This comparison suggests that near $x=0.25$, $T_c$ at the physical $J_{\perp}/J_{\parallel}=1.75$ is more than an order of magnitude higher than that at $J_{\perp}=0$.
Therefore, under realistic parameters for LNO, the pairing state manifests as interlayer $s$-wave pairing,
highlighting the robust inter-layer pairing  driven by enhanced inter-layer superexchange.

\begin{figure}[t!]
\centering
\includegraphics[width=0.95\textwidth]{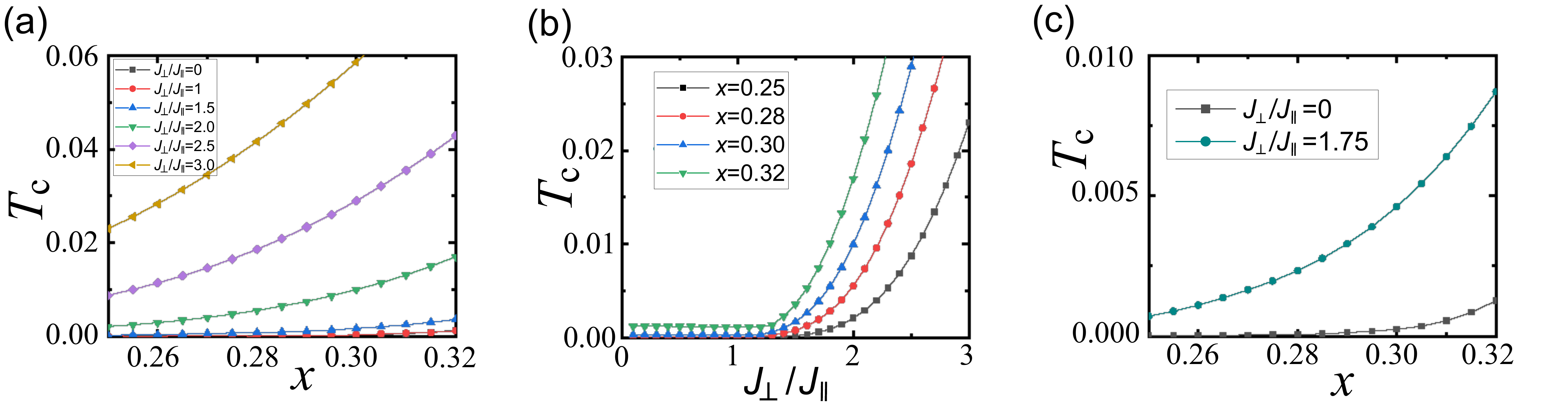}
\caption{(a) Superconducting $T_c$ versus filling $x$ at $J_{\parallel}=0.4t$ for different coupling ratios $J_{\perp}/J_{\parallel}=0, 1, 1.5, 2, 2.5, 3$.
(b) $T_c$ versus $J_{\perp}/J_{\parallel}$ for different filling $x=0.25, 0.28, 0.3, 0.32$.
(c) Comparison of $T_c$ versus filling $x$ between $J_{\perp}/J_{\parallel}=0$ and $J_{\perp}/J_{\parallel}=1.75$. }
\label{fig:JperpTcFilling}
\end{figure}

In summary, our findings regarding the dependence of $T_c$ on $J_{\perp}/J_{\parallel}$ and $x$ are consistent with experimental observations in LNO under pressure.
The high-$T_c$ superconductivity in LNO, observed under pressure \cite{Wang2023LNO}, can be explained by the significant enhancement of the superconducting pairing arising from increased interlayer coupling.
The theory predicts that electron doping enhances $T_c$ effectively, whereas hole doping suppresses it, providing insights into the doping-dependent behavior of superconductivity in LNO under pressure.

\paragraph{Discussion:}
For comprehensiveness, let us discuss the feasibility of SBMFT.
In our study, a two-dimensional (2D) bilayer of unit-cell is extracted from the quasi-2D layered structure of LNO.
The superconducting critical temperature $T_c$, derived through SBMFT for such 2D single bilayer system, primarily signifies the temperature $T_{\text{pair}}$ at which pairs form,
albeit the actual onset of superconductivity necessitates long-range phase coherence.
For the quasi-2D superconductor, the weak inter-bilayer coupling, which always exists, facilitates the establishment of long-range phase coherence essential for superconductivity.
Even a tiny inter-bilayer coupling proves sufficient for initiating long-range phase coherence \cite{kopec2000}.
Notably, further enhancing the inter-bilayer coupling does not significantly enhance $T_c$.
This scenario also applies to our case that $T_c$ should not be sensitive to the inter-bilayer coupling on condition that it has already set up long-range phase coherence.

On the other hand, the interlayer coupling within the bilayer is crucial for the superconducting pair formation.
The strong interlayer super-exchange interaction $J_{\perp}$ induces a robust interlayer $s$-wave pairing,
and significantly enhances $T_c$.
Indeed, the high-$T_c$ superconductivity in LNO only emerges when the imposed pressure exceeds a critical value, i.e. $14$GPa, upon which the structure transition takes place \cite{Wang2023LNO}.
This experimental observation suggests that the enhancement of the interlayer coupling brought about by the structure transition is crucial for the onset of high-$T_c$ superconductivity in pressurized LNO.
The mean-field $T_c$ obtained in this work is meaningful because an additional weak inter-bilayer coupling,
which always exists in the real material, will stabilize the long-range superconducting order \cite{kopec2000}.

\end{widetext}

\end{document}